\documentclass[authoryear,a4paper,11pt]{elsarticle_WP}

\newif\ifdetails
\detailsfalse

\usepackage{hyperref}
\usepackage{graphicx}
\usepackage{amssymb,amsmath}
\usepackage{bm}
\usepackage{bbm}
\usepackage[margin=1in]{geometry}
\usepackage{color}
\usepackage{natbib}
\usepackage{setspace}
\usepackage{multirow}
\usepackage{booktabs}
\usepackage{array}
\usepackage[normalem]{ulem}
\usepackage{tikz}
\usepackage{endnotes}

\let\footnote=\endnote

 \newcolumntype{x}[1]{>{\centering\arraybackslash\hspace{0pt}}p{#1}}

\journal{}

\newtheorem{proposition}{Proposition}

\newcommand{\zerob} {{\bf 0}}
\newcommand{\oneb} {{\bf 1}}

\newcommand{\thetab} {{\boldsymbol{\theta}}}

\newcommand{\nub} {{\boldsymbol{\nub}}}

\newcommand{\Deltab} {{\boldsymbol{\Delta}}}

\newcommand{\Thetab} {{\boldsymbol{\Theta}}}

\newcommand{\intd} {\textrm{d}}

\newcommand{\zetab} {\boldsymbol{\zeta}}

\newcommand{\Sigmamat} {{\bm \Sigma}}

\newcommand{\Psib} {{\bm \Psi}}

\newcommand{\Amat} {\textbf{A}}
\newcommand{\Bmat} {\textbf{B}}

\newcommand{\Dvec} {\textbf{D}}
\newcommand{\Gmat} {\textbf{G}}

\newcommand{\Qmat} {\textbf{Q}}
\newcommand{\Rmat} {\textbf{R}}

\newcommand{\Cmat} {\mathbf{C}}

\newcommand{\im} {\iota}

\newcommand{\Xmat} {\textbf{X}}
\newcommand{\Xvec} {\mathbf{X}}
\newcommand{\Uvec} {\mathbf{U}}
\newcommand{\Rvec} {\mathbf{R}}
\newcommand{\Imat} {\textbf{I}}

\newcommand{\Vmat} {\textbf{V}}

\newcommand{\avec} {\textbf{a}}

\newcommand{\hvec} {\textbf{h}}
\newcommand{\xvec} {\textbf{x}}

\newcommand{\svec} {\textbf{s}}
\newcommand{\tvec} {\textbf{t}}
\newcommand{\uvec} {\textbf{u}}

\newcommand{\muvec} {\boldsymbol{\mu}}

\newcommand{\betab} {\boldsymbol {\beta}}

\renewcommand{\zerob}{\mathbf{0}}

\newcommand{\x}{\mathbf{x}}

\newcommand{\Yvec}{\mathbf{Y}}
\newcommand{\Wvec}{\mathbf{W}}
\newcommand{\Gvec}{\mathbf{G}}

\newcommand{\Zvec}{\mathbf{Z}}

\newcommand{\E}{\mathrm{E}}
\newcommand{\cov}{\mathrm{cov}}
\newcommand{\var}{\mathrm{var}}

\newcommand{\bSigma}{\bm{\Sigma}}

\let\originalleft\left
\let\originalright\right
\renewcommand{\left}{\mathopen{}\mathclose\bgroup\originalleft}
\renewcommand{\right}{\aftergroup\egroup\originalright}

\begin{document}

\begin{frontmatter}

\title{Non-Gaussian bivariate modelling with application to atmospheric trace-gas inversion}

% \author{Andrew Zammit Mangion\footnote{National Institute for Applied Statistics Research Australia~(NIASRA), University of Wollongong, Australia} \and Noel Cressie\footnotemark[1] \and Anita L. Ganesan\footnote{School of Geographical Sciences, University of Bristol, UK} \and Matthew Rigby\footnote{Department of Chemistry, University of Bristol, UK}}
% \date{}

\author[Wollongong]{Andrew Zammit-Mangion\corref{correspondingauthor}}
\cortext[correspondingauthor]{Corresponding author}
\ead{azm@uow.edu.au}
\author[Wollongong]{Noel Cressie}
\ead{ncressie@uow.edu.au}
\author[Bristol]{Anita L. Ganesan}
\ead{anita.ganesan@bristol.ac.uk}
\address[Wollongong]{National Institute for Applied Statistics Research Australia~(NIASRA), School of Mathematics and Applied Statistics (SMAS), University of Wollongong, Northfields Avenue, Wollongong, NSW 2522, Australia}
\address[Bristol]{School of Geographical Sciences, University of Bristol, University Road, Bristol, BS8 1SS, UK}

\begin{abstract}

Atmospheric trace-gas inversion is the procedure by which the sources and sinks of a trace gas are identified from observations of its mole fraction at isolated locations in space and time. This is inherently a spatio-temporal bivariate inversion problem, since the mole-fraction field evolves in space and time and the flux is also spatio-temporally distributed. Further, the bivariate model is likely to be non-Gaussian since the flux field is rarely Gaussian. Here, we use conditioning to construct a non-Gaussian bivariate model, and we describe some of its properties through auto- and cross-cumulant functions. A bivariate non-Gaussian, specifically trans-Gaussian, model is then achieved through the use of Box--Cox transformations, and we facilitate Bayesian inference by approximating the likelihood in a hierarchical framework. Trace-gas inversion, especially at high spatial resolution, is frequently highly sensitive to prior specification. Therefore, unlike conventional approaches, we assimilate trace-gas inventory information with the observational data at the parameter layer, thus shifting prior sensitivity from the inventory itself to its spatial characteristics (e.g., its spatial length scale). We demonstrate the approach in controlled-experiment studies of methane inversion, using fluxes extracted from inventories of the UK and Ireland and of Northern Australia.

\end{abstract}


\end{frontmatter}

%\linenumbers

\section{Introduction}\label{sec:Intro}

Atmospheric trace-gas inversion is the procedure by which flux fields (gas sinks and sources) are identified from gas mole-fraction observations. Unlike conventional problems in spatial statistics, the spatial field of principal interest, the flux field, is rarely directly observed. Instead, satellite and surface gas-concentration instruments are sensitive to gas particles after they have been transported over potentially large distances over time. Spatio-temporal statistical methodology is well placed to obtain global space-time maps of mole fractions of an atmospheric constituent from irregular observations \citep[see, for example,][]{Cressie_2010, Cameletti_2013, Lindgren_2011}; however, the inversion of these maps in order to pinpoint the sources and sinks of the gas is a much more difficult problem. Its solution is key to effective policy implementation with regard to greenhouse gas emissions and climate change \citep{IPCC_2014}.

Current approaches to trace-gas inversion build on the data-assimilation framework described, for instance, in \cite{Tarantola_2005}, where prior beliefs on a flux field are updated with observations, to produce posterior beliefs \citep[see, for example,][]{Rigby_2011,Stohl_2009}. The prior distribution is usually formulated to have prior expectation equal to values in an \emph{inventory}, a flux database constructed from auxiliary activity data (e.g., vehicles per unit area) and emission factors associated with the emission sector (e.g., carbon dioxide emissions per vehicle), while the prior covariance is usually constructed using values calculated from the inventory. Reasonable prior marginal variances are typically assumed; for example, a prior density function at each location may be chosen such that the area under the curve between 0.5 and 1.5 times the value of the inventory at that location is around 68\% \citep{Ganesan_2014}. Frequently, a diagonal structure is imposed on the prior covariance for all locations. It is also usually assumed that the prior expectation and prior covariance completely specify the prior distribution.

These current approaches can be critiqued due to their use of inappropriate or inflexible models. First, atmospheric trace-gas inversion is inherently a bivariate spatio-temporal problem, where observations are not readings of fluxes, but rather readings of a second, mole-fraction, field. The mole-fraction field is generated by the underlying flux field and by meteorology, which determines transport of the trace gas. Making this distinction allows one to attribute uncertainties appropriately, either to instrumentation error or to imprecise mole-fraction field modelling (e.g., due to linearisation of the flux-mole-fraction mapping or imprecise specification of transport modelling/boundary conditions). Second, a Gaussian model is often inappropriate for the flux field \citep{Ganesan_2014,Miller_2014}, which can be inherently non-negative and which, as seen from inventories, may exhibit skewness and higher-order features. Third, spatial correlations in prior error covariances, even when using regressors, should be assumed since errors in the inventories are likely to be spatially correlated. Fourth, several works claim that inventories are highly inaccurate in certain regions \citep[e.g.,][]{Lunt_2015}. Consequently, there is a drive to divert from use of the inventories in the model, and to take a data-driven approach to flux inversion, even when sub-national resolution is required and so prior information is especially important. 

These reservations were first discussed in \cite{Zammit_2015}, where the authors constructed a hierarchical, lognormal bivariate spatio-temporal model for flux inversion of methane in the UK and Ireland. In that article, an empirical hierarchical modelling approach \citep[][Section 2.1]{Cressie_2011} was adopted, where numerous parameters were estimated offline prior to drawing samples from the empirical posterior distribution of the flux field. In this article, we extend the modelling and inferential approach in \cite{Zammit_2015} in two ways. First, we relax the assumption of lognormality by defining a more general non-Gaussian model, and subsequently we model the flux field as a trans-Gaussian (here a Box--Cox) spatial process \citep{deOliveira_1997}. The class of Box--Cox spatial processes includes the lognormal spatial process as a special case. Second, following a likelihood approximation, we adopt a fully Bayesian approach to flux-field inversion that naturally propagates variability in the parameters \citep[parameters that were estimated offline in][]{Zammit_2015} to our inferences on the flux field. In order to reduce reliance on the flux inventory, which is based on emission factors that are not precisely known, we do not assume it to be the prior mean of the flux field. Instead, we take a new approach by assuming that it is an independent realisation of the flux field that we wish to infer. Hence, we assume that the inventory is informative of the spatial (prior) properties of the flux field, but not of the flux field itself, so that posterior inferences on fluxes are still predominantly data-driven. We demonstrate the value of this approach in a controlled experiment where we infer flux fields in the UK using the true flux as our inventory, an inventory with similar spatial properties (extracted from mainland Europe), and one with highly dissimilar spatial properties (from Northern Australia). In the experiment, inferred flux fields are compared to the known, true flux field.

The article is organised as follows. In Section \ref{sec:overview} we motivate atmospheric trace-gas inversion as a problem in bivariate modelling \citep{Cressie_2015,Royle_1999} and outline a few of the proposed bivariate model's theoretical properties. In Section \ref{sec:hierarchical} we focus on the bivariate trans-Gaussian model and place it within a hierarchical framework appropriate for describing the problem of trace-gas inversion. The transformation that we feature here is the well known Box--Cox transformation. The hierarchical model we employ is summarised in Section \ref{sec:summary}; the succinct summary presented there can help to put the model descriptions of the preceding sections into perspective. In Section \ref{sec:inference} we outline the Markov chain Monte Carlo (MCMC) scheme adopted and the approximations we use to facilitate its implementation. In Section \ref{sec:semi-empirical} we apply the framework to a realistic simulation study of methane emissions in the UK and Ireland, where fluxes and synthetic mole-fraction observations are simulated using meteorology from a numerical model and UK/Ireland flux inventories. To show the flexibility of the model, we also apply it to the case when mole fractions are simulated from a Northern Australian inventory. The methane emissions in this region are considerably smoother (spatially) than those in the UK and Ireland. Conclusions are drawn in Section \ref{sec:conc}, and the article finishes with three technical appendices.

\section{Application and modelling overview}\label{sec:overview}

In this section we motivate the problem of atmospheric trace-gas inversion as a problem in non-Gaussian bivariate modelling (Section \ref{sec:inversion_intro}), and discuss some properties of the non-Gaussian bivariate model (Section \ref{sec:nonGaussian}).

\subsection{Atmospheric trace-gas inversion}\label{sec:inversion_intro}

Monitoring of greenhouse gases and air pollutants is a key priority for environmental agencies and institutions worldwide. One of its primary aims is to be able to determine where the gas is being produced (sources) and where it is being removed (sinks). This is not a straightforward task, since the net flux of the production and removal processes is related to the observed mole fraction at a particular time and location after transport by meteorology and alteration by chemical reactions. These processes can only be obtained from computationally intensive numerical models.

The numerical models employed are either Eulerian \citep[e.g., the Comprehensive Air quality Model with eXtensions (CAMx); see][]{Emery_2012} or Lagrangian \citep[e.g., the Numerical Atmospheric-dispersion Modelling Environment (NAME) or the FLEXible PARTicle dispersion model (FLEXPART); see][]{Jones_2007,Thompson_2014} in nature. Although these two classes of models are fundamentally different, they serve the same purpose in inversion, namely to establish the sensitivity of the mole fractions to the flux. In the application we consider in Section \ref{sec:semi-empirical}, we shall use a Lagrangian model for regional inversion (specifically the model NAME), however the inversion framework we construct is applicable to both classes of models.

Now, consider a spatially referenced flux field, $Y_1(\cdot)$, and a spatially referenced mole-fraction field (at some given time instant), $Y_2(\cdot)$. Then meteorology and loss (chemical) processes imply that 
\begin{equation}\label{eq:model1}
Y_2(\cdot) = \mathcal{H}(Y_1(\cdot)),
\end{equation}
where $\mathcal{H}(\cdot)$ is some nonlinear operator that Eulerian or Lagrangian models attempt to reconstruct. Following \citet[][Section 7.3]{Cressie_2011}, a stochastic version of this model can be written, conditional on $Y_1(\cdot)$, as
\begin{equation}\label{eq:model2}
Y_2(\cdot) = \mathcal{H}(Y_1(\cdot); \xi(\cdot)),
\end{equation}
\noindent where $\xi(\cdot)$ represents a process that expresses uncertainty in the physical relationship \eqref{eq:model1} that defines a model for $Y_2(\cdot)$ in terms of $Y_1(\cdot)$.

Consequently, \eqref{eq:model2} defines a bivariate stochastic process that incorporates not only knowledge about inventories and transport, but also uncertainty around how that knowledge results in the mole-fraction field $Y_2(\cdot)$.  With a few exceptions \citep[e.g.,][]{Peters_2005}, a linear mapping that maintains the physical relationship to leading order, $\mathcal{H}_L$, is substituted for $\mathcal{H}$, and a statistical model
\begin{equation}\label{eq:model3}
Y_2(\cdot) = \mathcal{H}_L(Y_1(\cdot)) + \zeta(\cdot), 
\end{equation}
is assumed, conditional on $Y_1(\cdot)$. In \eqref{eq:model3}, $\zeta(\cdot)$ collects discrepancies and uncertainties from (i) using leading-order, linear physical relationships and (ii) inaccuracies inherent to the chemical transport model. When analysing short-lived gases, such as carbon monoxide, the sink due to chemical reaction (a component of $\mathcal{H}$) can be nonlinearly dependent on the flux. However, for long-lived gases such as methane,  considered in Section \ref{sec:semi-empirical}, this sink is approximately flux-independent. We therefore expect uncertainties arising from (i) to be dominated by (ii). When using $\mathcal{H}_L$ instead of $\mathcal{H}$, model interpretation and inference is greatly facilitated. In all that follows, we base our work on \eqref{eq:model3}.

Using this conditional approach to building a bivariate process that respects the physical system expressed (to leading order) by \eqref{eq:model3}, is very powerful, since it now only remains to define a univariate spatial process for $Y_1(\cdot)$. That process need not be Gaussian, and indeed for the methane flux fields considered in Section \ref{sec:semi-empirical}, it will generally not be. Consequently, from \eqref{eq:model2} and \eqref{eq:model3} the joint process $(Y_1(\cdot),Y_2(\cdot))$ is tied to both the physics expressed in $\mathcal{H}_L$ and the stochastic (possibly non-Gaussian) properties of $Y_1(\cdot)$.

 The joint statistical properties of $Y_1(\cdot)$ and $\zeta(\cdot)$ fully determine the statistical properties of $Y_2(\cdot)$. We briefly discuss these properties for general $Y_1(\cdot)$ and  when $\mathcal{H}_L$ is an integral transform in Section \ref{sec:nonGaussian}. Unlike $Y_2(\cdot)$, observations of $Y_1(\cdot)$, when available, only capture the process' very fine scales. Hence one cannot enjoy the use of typical exploratory-data-analysis techniques to guide the choice of model for $Y_1(\cdot)$. The presence of \emph{flux inventories} can help, although over-reliance on these inventories, which are hard to validate, is an ongoing concern \citep[e.g.,][]{Berchet_2015}; we return to this point in Section \ref{sec:inventory}.

This article is concerned with an appropriate model choice for $Y_1(\cdot)$ and $\zeta(\cdot)$. Both for analytical and computational convenience, $Y_1(\cdot)$ and $\zeta(\cdot)$ are frequently assumed to be Gaussian \citep[e.g.,][]{Stohl_2009} in which case $Y_2(\cdot)$ is Gaussian and $(Y_1(\cdot),Y_2(\cdot))$ is jointly Gaussian. In recent years, there has been a shift away from Gaussianity assumptions; for example, \citet{Rigby_2011} considered exponential distributions for $Y_1(\cdot)$ evaluated at the grid-cell level, \citet{Ganesan_2014} considered lognormal distributions, and  \citet{Miller_2014} considered a truncated spatial Gaussian process for $Y_1(\cdot)$. \citet{Zammit_2015} showed that both lognormality and spatial correlation in $Y_1(\cdot)$ are important for prediction in the UK and Ireland. However, it is also the case that lognormality may not be a suitable assumption everywhere and at all grid resolutions. A flexible class of models that can accommodate different degrees of non-Gaussianity and spatial correlation is the class of trans-Gaussian spatial processes \citep[][p.~137]{Cressie_1993}; we use them within a trace-gas inversion framework in Section \ref{sec:hierarchical} and illustrate the importance of the added flexibility in Section \ref{sec:semi-empirical}. 

\subsection{Properties of the bivariate non-Gaussian model}\label{sec:nonGaussian}

If $Y_1(\cdot)$ and $\zeta(\cdot)$ are Gaussian processes then the properties of the joint process $(Y_1(\cdot),Y_2(\cdot))$ are straightforward to derive \citep{Cressie_2015}. When $Y_1(\cdot)$ is a lognormal process, then the first-order and second-order moments of the finite-dimensional distributions of $(Y_1(\cdot),Y_2(\cdot))$ can be expressed in closed form \citep{Zammit_2015}. In this section we discuss the properties of the joint  process $(Y_1(\cdot),Y_2(\cdot))$ when $Y_1(\cdot)$ is any (in general, non-Gaussian) process.

Consider a real-valued random field $\{Y_1(\svec): \svec \in D \subset \mathbb{R}^d\}$, where typically $d = 2$ and $D$ is a continuously indexed domain. Construct the real-valued field $\{Y_2(\svec): \svec \in D \subset \mathbb{R}^d\}$ through
\begin{equation}\label{eq:Y2}
Y_2(\svec) = \int_D b(\svec,\uvec)Y_1(\uvec) \intd \uvec + \zeta(\svec); \quad \svec \in D,
\end{equation}
where we term the integral kernel $b(\cdot,\cdot)$ an interaction function and $\zeta(\cdot)$ is a discrepancy term independent of $Y_1(\cdot)$. 
%Recall that in trace-gas inversion the discrepancy $\zeta(\cdot)$ captures variability in $Y_2(\cdot)$ that is not explained by the integral transform of $Y_1(\cdot)$,  % For computational reasons made apparent in Section \ref{sec:hierarchical}, we assume that $\zeta(\cdot)$ is a Gaussian process; we further assume that $\E(\zeta(\cdot)) = 0$. 
%$Y_2(\cdot)$  denotes the mole-fraction field at a single time instance, $Y_1(\cdot)$ a time-averaged flux field, and $b(\cdot,\cdot)$ the (assumed known) meteorology-driven relationship that relates the two at this time instance \citep{Zammit_2015}. %The flux field $Y_1(\cdot)$ is non-Gaussian; hence, whilst $Y_2(\cdot)$ is conditionally (given $Y_1(\cdot)$) Gaussian, it is unconditionally also non-Gaussian. 
The integral in \eqref{eq:Y2} exists for all $\svec \in D$ as a quadratic mean limit of Riemann sums provided that
\begin{equation}\label{eq:ddint}
C_{Y_2Y_2}(\svec_1,\svec_2) \equiv \int_D\int_Db(\svec_1,\uvec_1)b(\svec_2,\uvec_2)C_{Y_1Y_1}(\uvec_1,\uvec_2)\intd \uvec_1 \intd \uvec_2 < \infty; \quad \svec_1, \svec_2 \in D,
\end{equation}
and $C_{Y_1Y_1}(\cdot,\cdot) \equiv \cov(Y_1(\cdot),Y_1(\cdot))$ is continuous on $D \times D$ (see \citet{Yaglom_1987}, p.~68, and \citet[][]{Lindgren_2012}, Theorem 2.16).\ifdetails\footnote{A sequence of random variables $\{X_n\}$ converges in quadratic mean to $X$ ($X_n \xrightarrow{q.m.} X$) if $\E[(X_n - X)^2] \rightarrow 0$ when $n \rightarrow \infty$. Convergence can be assessed using either the \emph{Cauchy} criterion or the \emph{Lo{\`e}ve} criterion \citep[][p.144]{Lindgren_2013}:
\begin{itemize}
\item Cauchy criterion: $\E[(X_m - X_n)^2] \rightarrow 0$,
\item Lo{\`e}ve criterion: $\E(X_mX_n) \rightarrow c$, a constant,
\end{itemize}
when $m,n \rightarrow \infty$ independently of each other.

Why convergence in quadratic mean? It can be shown \citep[][p.62]{Yaglom_1987}, through Chebyshev's inequality, that convergence in quadratic mean is actually a stronger condition that what one would naturally take as the definition of a limit, that is,
$$P(|X - X_n| > \epsilon) = 0, \quad n \rightarrow \infty,$$
and that convergence in quadratic mean implies a \emph{limit in probability} of the sequence. It is also true (p.64) that if the Cauchy criterion is met, $X_n \rightarrow X$ as $n\rightarrow \infty$.}\fi\ifdetails$^,$\footnote{Theorem 2.16 in \citet{Lindgren_2012} states that if the covariance function of a stochastic process $x(t)$, $r(s,t)$, is continuous in $[a,b] \times [a,b]$, $\E(x(t)) = 0$, and $g(t)$ is such that the Riemann integral $$Q_1 \equiv \int_{a}^b\int_a^bg(s)\overline{g(t)}r(s,t)\intd s \intd t < \infty,$$ then $J_1 \equiv \int_a^b g(t)x(t)\intd t$ exists as a quadratic mean limit of a Riemann sum and $\E(J_1) = 0$ (if $\E(x(t)) = 0$) and $\E(|J_1|^2) = Q_1$. The proof follows the Lo{\`e}ve criterion. Let $S_m \equiv \sum_{k=1}^m g(t_k)x(t_k)(t_k - t_{k-1})$. Then$$\E(S_m\overline{S_n}) = \sum_{k=1}^m\sum_{j=1}^n g(s_k)\overline{g(t_j)}r(s_k,t_j)(s_k - s_{k-1})(t_j - t_{j-1}),$$ which converges to a finite quantity, $Q_1 < \infty$, as $m,n \rightarrow \infty$. Therefore, $\E[(S_n - J_1)^2] \rightarrow 0$ as $n \rightarrow \infty$.}
\fi~Thus, provided that $\zeta(\cdot)$ is well defined and \eqref{eq:ddint} holds, $Y_2(\cdot)$ exists by construction. Further, from \eqref{eq:ddint}, if $\frac{\partial}{\partial \svec}b(\svec,\cdot)$ exists for all $\svec \in D$, and if $\zeta(\cdot)$ is differentiable in quadratic mean, then $Y_2(\cdot)$ is differentiable in quadratic mean as well \citep[][p.37]{Aastrom_2006}.\ifdetails\footnote{
This statement is a bit different in \citet[][p.~153]{Zhong_2004}, who implies that if  $Y_1(\cdot)$ is mean square integrable and if $b(\svec,\uvec)$ is continuously integrable, then if $\zeta(\svec)$ is mean square differentiable, $Y_2(\cdot)$ is as well. I think it is differentiability of $b(\svec,\uvec)$ which is important. 

A proof for differentiability (in the nonstationary) case is provided by \citet{Aastrom_2006}, p.37. His Theorem 6.4 states that a second order stochastic process $\{x(t), t \in T\}$ is differentiable in mean square at $t_0 \in T$ if and only if the mean value function $m(t)$ is differentiable at $t_0$ and the generalised second order derivative of the covariance function, $$\frac{\partial^2r(s,t)}{\partial{s}\partial{t}},$$ exists at $s = t = t_0$. The proof relies once again on the Lo{\`e}ve criterion. It is easy to see that for the conditional approach, differentiability in the mean and covariance of $Y_2(\cdot)$ reduces to the condition that $b(\svec,\cdot)$ is differentiable for all $\svec$, since, for $\zeta(\svec) = 0$,

$$ \frac{\partial C_{22}(\svec_1,\svec_2)}{\partial \svec_1 \partial \svec_2} = \int_D\int_D \frac{\partial}{\partial\svec_1}b(\svec_1,\uvec)\frac{\partial}{\partial\svec_2}b(\svec_2,\uvec)C_{11}(\uvec_1,\uvec_2)\intd\uvec_1\intd\uvec_2,$$
and
$$\frac{\partial \E(Y_2(\svec))}{\partial \svec} = \int_D \frac{\partial}{\partial\svec}b(\svec,\uvec)\E(Y_1(\uvec))\intd \uvec.$$

 Interestingly, differentiability of $Y_2(\cdot)$ does not depend on differentiability of $Y_1(\cdot)$.}\fi

%%##See aslso Wirschang A.5 (Vibrations)

%Gaussian bivariate models constructed through an integral transform are flexible, and easily amenable to represent characteristics such as spatial heterogeneity and asymmetry in the behaviour of the joint process $(Y_1(\cdot),Y_2(\cdot))$ which we can expect in this application; see \cite{Cressie_2015}. When $Y_1(\cdot)$ is non-Gaussian, the joint model enjoys this same flexibility with the added benefit that characteristics associated with third-order and higher-order moments can also be modelled. Now, due to the form of \eqref{eq:Y2}, it turns out that one can succintly express the properties of the joint process $(Y_1(\cdot),Y_2(\cdot))$ in terms of the cumulants of the finite-dimensional distributions of $Y_1(\cdot)$ and $\zeta(\cdot)$.
% and the additive property of cumulants for independent random variables \citep[e.g.,][Theorem 4.16]{Severini_2005},
\ifdetails\footnote{For a random variable $X$, the $j$-th cumulant, $\kappa_j$, is found through the logarithm of its characteristic function $\varphi_X(t)$ as follows \citep[][p.114]{Severini_2005}:
$$ \kappa_j =\left. \frac{1}{(\im)^j} \frac{\intd^j}{\intd t^j} \ln \varphi_X(t)\right|_{t=0}.$$
A similar expression can be obtained relating the cumulant to the moment generating function (MGF), however the MGF does not always exist. The expression for a multivariate variable $\Xvec$ is similar, although now one usually denotes the \emph{order} of a cumulant as a sequence of subscripts. For example, 
$$ \kappa_{i_1,\dots,i_d} =\left. \frac{1}{(\im)^n}\frac{\partial^{i_1 + \cdots +i_d}}{\partial t_1^{i_1}\cdots \partial t_d^{i_d}} \ln \varphi_\Xvec({\bf t})\right|_{{\bf t} =\zerob}, \quad n = \sum_{j=1}^d i_j.$$
In this article I have chosen the subscript to reference the process, while the superscript to reference the order. Implicitly, a superscript of $2$ indicates an order of $(1,1)$. A superscript of $3$ indicates an order of $(1,1,1)$ and so on.}\fi\ifdetails$^,$\footnote{The additive property of cumulants is as follows. Let $\Xvec,\Yvec$ denote two $d$-dimensional random variables such that all cumulants of $\Xvec$ and $\Yvec$ exist, and assume that $\Xvec$ and $\Yvec$ are independent. Let $\kappa_{i_1\cdots i_d}(\Xvec), \kappa_{i_1\cdots i_d}(\Yvec),$ and $\kappa_{i_1\cdots i_d}(\Xvec + \Yvec)$ denote the cumulant of order $(i_1,\dots,i_d)$ of $\Xvec, \Yvec$ and $\Xvec + \Yvec$, respectively. Then $$ \kappa_{i_1,\cdots i_d}(\Xvec + \Yvec) = \kappa_{i_1,\cdots i_d}(\Xvec) + \kappa_{i_1,\cdots i_d}(\Yvec).$$ In our application the transformed $Y_1(\cdot)$ and the discrepancy $\zeta(\cdot)$ in \eqref{eq:Y2} are independent and therefore we can take advantage of this property.}  \fi%the properties of $Y_2(\cdot)$ and, in general, the joint process $(Y_1(\cdot),Y_2(\cdot))$, are best illustrated in terms of the cumulants of $Y_1(\cdot)$ and $\zeta(\cdot)$ evaluated at a finite set of locations. 

%Description of stochastic processes in terms of spatial cumulant functions has a long history in signal processing \cite[e.g.,][]{Mendel_1991} but has received less attention in spatial statistics \citep{Dimitrakopoulos_2010}.

Without loss of generality, we consider the (joint) spatial cumulants of the processes evaluated at $n \ge 1$, possibly repeated, locations. Let $\varphi_{\Yvec_1}(\tvec_1)$ be the characteristic function of the vector $(Y_1(\uvec_1),\dots,Y_1(\uvec_n))'$, where it is understood that $\varphi_{\Yvec_1}(\tvec_1)$ is a function of the locations $\uvec_1,\dots,\uvec_n$. We define the $n$-th order \emph{spatial cumulant function} of $\uvec_1,\dots,\uvec_n$ to be
\begin{equation}\label{eq:autocumulant}
\kappa_{Y_1\dots Y_1}^n(\uvec_1,\dots,\uvec_n) \equiv \frac{1}{\iota^n}\left.\frac{\partial^n}{\partial t_{11}\dots\partial t_{1n}}\ln \varphi_{\Yvec_1}(\tvec_1)\right|_{\tvec_1 = \zerob},
\end{equation}
\noindent where the superscript of $\kappa(\cdot)$ denotes the spatial cumulant order, the subscript denotes the processes associated with the spatial locations, and $\iota$ is the imaginary unit. Since \eqref{eq:autocumulant} is dependent only on $Y_1$, it is an \emph{auto-cumulant function}. \emph{Cross-cumulant functions} can be constructed by assuming that a possibly different process is evaluated at each location, in which case
\begin{equation}\label{eq:crosscumulant}
\kappa_{Y_{j_1}\dots Y_{j_n}}^n(\svec_1,\dots,\svec_n) \equiv \frac{1}{\iota^n}\left.\frac{\partial^n}{\partial t_{{j_1}1}\dots\partial t_{{j_n}n}}\ln \varphi_{\Yvec}(\tvec)\right|_{\tvec = \zerob},
\end{equation}
where $j_k \in \{1,2\},$ for $k = 1,\dots,n$; $\Yvec = (Y_{j_k}(\svec_k): k = 1,\dots,n)'$; and $\tvec = (t_{{j_k}k} : k = 1,\dots,n)'$.

%If the subscript shows identical processes, then $\kappa(\cdot)$ is an \emph{auto-cumulant function}, otherwise it is a \emph{cross-cumulant function}.

We are interested in characterising the joint process $(Y_1(\cdot),Y_2(\cdot))$ when the auto-cumulant functions associated with $Y_1(\cdot)$ are known. We start by considering the auto-cumulant functions of $Y_2(\cdot)$. Since the cumulant of a sum of independent random variables is the sum of the individual cumulants, the auto-cumulant functions associated with $Y_2(\cdot)$ are a sum of those associated with the processes, $\int_D b(\cdot,\uvec)Y_1(\uvec)\intd \uvec$ and $\zeta(\cdot)$. Hence, it can be shown through use of the characteristic function \citep[][Appendix III]{Kuznetsov_1965} that, provided all auto-cumulant functions associated with $Y_1(\cdot)$ are integrable,
% Denote the first, second, third, and $n$-th order cumulants (first order in each variable) associated with $Y_1(\cdot)$ at $\uvec_1,\dots,\uvec_n$ as $$\kappa_{Y_1}^1(\uvec_1), \kappa_{Y_1Y_1}^2(\uvec_1,\uvec_2),\kappa_{Y_1Y_1Y_1}^3(\uvec_1,\uvec_2,\uvec_3),\dots,\kappa^n_{Y_1\dots Y_1}(\uvec_1,\dots,\uvec_n),$$ respectively, where $\uvec_1,\dots,\uvec_n \in D$. Here, the superscript indicates the combined cumulant order and the subscript the processes associated with each respective spatial location (each subscript can be either `$Y_1$', `$Y_2$' or `$\zeta$'). Assume that all cumulants associated with $Y_1(\cdot)$ are integrable. By taking a fine grid over $D$ to represent the integral in \eqref{eq:Y2} and letting the grid spacing tend to zero \citep[][Appendix III]{Kuznetsov_1965} and by taking advantage of the additive property of cumulants, we obtain
 \begin{align}\label{eq:12cumulants}
 \kappa_{Y_2}^1(\svec_1) &= \int_D b(\svec_1,\uvec_1)\kappa_{Y_1}^1(\uvec_1) \intd \uvec_1 + \kappa_\zeta^1(\svec_1); \quad \svec_1 \in D,\\
 \kappa_{Y_2Y_2}^2(\svec_1,\svec_2) &= \int_D\int_D b(\svec_1,\uvec_1)b(\svec_2,\uvec_2)\kappa_{Y_1Y_1}^2(\uvec_1,\uvec_2) \intd \uvec_1 \intd\uvec_2  + \kappa_{\zeta\zeta}^2(\svec_1,\svec_2);\quad \svec_1,\svec_2 \in D, \label{eq:12cumulantsb}
 \end{align}
and the $n$-th order auto-cumulant function associated with $Y_2(\cdot)$ is, for $n \ge 1$,
\ifdetails\footnote{
This follows from the fact that the terms in the exponent of $\varphi_{\Yvec_2}(\tvec_2)$ follow a regular pattern. Consider, for example, the third cumulant. The third term in the exponent of $\varphi_{\Yvec_1}(\tvec_1)$ is

$$\frac{\im^3}{3!}\sum_{j,j',j'' = 1}^{N,N,N} \kappa^3_{Y_1Y_1Y_1}(\uvec_j,\uvec_{j'},\uvec_{j''})t_{1j}t_{1j'}t_{1j''},$$

and therefore the third term in the exponent of $\varphi_{\Yvec_2}(\tvec_2)$ is 

$$ \frac{\im^2}{2}\sum_{j,j',j'' = 1}^{N,N,N}\sum_{i,i',i'' =1}^{M,M,M} t_{2i}t_{2i'}t_{2i''}b(\svec_i,\uvec_j)b(\svec_{i'},\uvec_{j'})b(\svec_{i''},\uvec_{j''})\kappa^3_{Y_1Y_1Y_1}(\uvec_j,\uvec_{j'},\uvec_{j''})\Delta_{\uvec_j}\Delta_{\uvec_{j'}}\Delta_{\uvec_{j''}}.$$

From the series expansion of $\varphi_{\Yvec_2}(\tvec_2)$ it follows that

$$ \kappa^3_{Y_2Y_2Y_2}(\svec_i,\svec_{i'},\svec_{i''}) = \sum_{j,j',j'' = 1}^{N,N,N} b(\svec_i,\uvec_j)b(\svec_{i'},\uvec_{j'})b(\svec_{i''},\uvec_{j''})\kappa^3_{Y_1Y_1Y_1}(\uvec_j,\uvec_{j'},\uvec_{j''})\Delta_{\uvec_j}\Delta_{\uvec_{j'}}\Delta_{\uvec_{j''}}.$$

\noindent The form of $\kappa^n$ is the same, irrespective of the order and, in general (Note: this result is not stated in \cite{Kuznetsov_1965} but is implied through the use of dots),

$$ \kappa^n_{Y_2Y_2Y_2}(\svec_i,\dots,\svec_{i^{(n)}}) = \sum_{j,\dots,j^{(n)} = 1}^N b(\svec_i,\uvec_j)\dots b(\svec_{i^{(n)}},\uvec_{j^{(n)}}) \kappa^n_{Y_1\dots Y_1}(\uvec_j,\dots,\uvec_{j^{(n)}})\Delta_{\uvec_j}\dots\Delta_{\uvec_{j^{(n)}}}.$$

}\fi
\begin{align}\label{eq:cumulants}
\kappa_{Y_2\dots Y_2}^n(\svec_1,\dots,\svec_n) = &\int_D\cdots\int_D b(\svec_1,\uvec_1)\dots b(\svec_n,\uvec_n) \kappa_{Y_1\dots Y_1}^n(\uvec_1,\dots, \uvec_n) \intd \uvec_1\dots\intd \uvec_n \nonumber \\ 
& + \kappa_{\zeta\dots \zeta}^n(\svec_1,\dots,\svec_n); \qquad\qquad \svec_1,\dots,\svec_n \in D.
\end{align}
%since the cumulants associated with a Gaussian process are zero for orders 3 and higher. 
When $\zeta(\cdot) = 0$ in \eqref{eq:Y2}, and hence $\kappa^n_{\zeta\dots\zeta}(\svec_1,\dots,\svec_n) \equiv 0$, we have
$$
\kappa_{Y_2\dots Y_2}^n(\svec_1,\dots,\svec_n) = \int_D\cdots\int_D b(\svec_1,\uvec_1)\dots b(\svec_n,\uvec_n) \kappa_{Y_1\dots Y_1}^n(\uvec_1,\dots, \uvec_n) \intd \uvec_1\dots\intd \uvec_n,
$$
which is a generalisation to cumulants of the bilinear property of covariances, namely $$\cov\left(\sum_{k=1}^{N_1} a_k U_{1k},\sum_{l=1}^{N_2} b_lU_{2l}\right) = \sum_{k=1}^{N_1}\sum_{l=1}^{N_2} a_k b_l \cov(U_{1k},U_{2l}),$$
where $\Uvec_1$ and $\Uvec_2$ are random vectors of length $N_1$ and $N_2$, respectively; $a_k \in \mathbb{R}$, for $k = 1,\dots,N_1$; and $b_l \in \mathbb{R}$, for $l = 1,\dots,N_2$.

In order to fully characterise the joint process $(Y_1(\cdot),Y_2(\cdot))$, we also need the cross-cumulant functions that involve both $Y_1(\cdot)$ and $Y_2(\cdot)$. This entails a lengthy but straightforward extension to the arguments of \citet{Kuznetsov_1965}, which we defer to \ref{sec:CCFs}. 
%In summary, in \ref{sec:CCFs}, we show that all cross-cumulant functions can be found by expressing the characteristic function of the finite-dimensional distributions of $(Y_1(\cdot),Y_2(\cdot))$, $\varphi_\Yvec(\tvec_\Yvec),$ in terms of that of the finite-dimensional distributions of $Y_1(\cdot)$, $\varphi_{\Yvec_1}(\tvec_1)$, and comparing this expression to a series expansion of $\varphi_\Yvec(\tvec_\Yvec)$. 
We show that the cross-cumulant functions are rather simple in form; for example, 
\begin{equation}\label{eq:cross-cum}
\kappa^3_{Y_1Y_1Y_2}(\uvec_1,\uvec_{2},\svec_{3}) = \int \kappa^3_{Y_1Y_1Y_1}(\uvec_1,\uvec_{2},\uvec_{3})b(\svec_{3},\uvec_{3})\intd\uvec_{3}.
\end{equation}
%Extensions to the arguments of \citet{Kuznetsov_1965} are needed to find the cross-cumulants, for example $\kappa_{Y_1Y_2}(\uvec_1,\svec_2)$. In \ref{sec:CCFs} we show how \emph{all} the cumulant functions associated with the joint process $(Y_1(\cdot),Y_2(\cdot))$ may be found in terms of those of $Y_1(\cdot)$ and $\zeta(\cdot)$ through the characteristic function.
All cross-cumulant functions of $(Y_1(\cdot),Y_2(\cdot))$ do not involve the process $\zeta(\cdot)$ since $Y_1(\cdot)$ and $\zeta(\cdot)$ are independent. Of all the auto-cumulant  and cross-cumulant functions, the second-order ones are arguably the most important since they are also the marginal and cross-covariance functions of the joint process. It can be shown \citep[][Section 3]{Cressie_2015} that the $2\times 2$ covariance matrix $(\kappa^2_{Y_iY_j}(\svec_1,\svec_2): i,j \in \{1,2\})$ given by \eqref{eq:12cumulantsb} is non-negative definite for all $\svec_1,\svec_2 \in D$.

% It can be shown \citep[][Section 3]{Cressie_2015} that the $2\times 2$ matrix $(\kappa_{Y_iY_j}(\svec_1,\svec_2): i,j \in \{1,2\})$, which is also a covariance matrix, is nonnegative definite by construction for all $\svec_1,\svec_2 \in D$. %Note that the cumulant functions were termed `generalized correlation functions' by \citet{Kuznetsov_1965}.

In the special case where $b(\svec,\uvec); \svec,\uvec \in D,$ is solely a function of $\svec - \uvec$, the spectra of the auto-cumulants of $Y_2(\cdot)$ (known as polyspectra)\ifdetails \footnote{\citet{Mendel_1991}.} \fi~can be found through a Fourier transform of \eqref{eq:cumulants}. However, in our application, $b(\cdot,\cdot)$ describes a sensitivity of mole fraction to flux at a given spatial location at a single time instance. It is meteorology-driven, and hence it is highly spatially dependent and asymmetric. Wind induces directional sensitivity and, consequently, the function $b(\svec,\cdot)$ frequently exhibits behaviour that is close to a step change at  $\svec$. As we show through a simple example in \ref{sec:cum_example}, this can result in interesting correlation structures and higher-order properties that are not possible to obtain using Gaussian, symmetric,  multivariate spatial models \citep[e.g., as found in ][]{Gneiting_2010}.

\section{Hierarchical trans-Gaussian bivariate models for atmospheric trace-gas inversion}\label{sec:hierarchical}

In this section we discuss the construction of our framework for atmospheric trace-gas inversion. First, in Section \ref{sec:hierarchical2} we develop the hierarchical model that uses the special class of non-Gaussian bivariate models, obtained by modelling $Y_1(\cdot)$ as a trans-Gaussian process with the Box--Cox spatial process as a leading case. In Section \ref{sec:inventory} we outline how to assimilate available flux inventories within the framework. In Section \ref{sec:summary}, we provide a succinct summary of the hierarchical model.

\subsection{The hierarchical model}\label{sec:hierarchical2}

In Section \ref{sec:nonGaussian} we considered the relationship of mole fraction to the flux field at a notional time instance. In practice, very little information on the flux field can be extracted from the mole-fraction field at a single time instance and, hence, considering its temporal evolution is important. In the following subsections we consider a spatio-temporal variant of the non-Gaussian model and place it within a Bayesian hierarchical framework appropriate for describing the problem of atmospheric trace-gas inversion.

\subsubsection{The mole-fraction observations}

Although the flux field $Y_1$ varies spatio-temporally, we treat it as a time-averaged spatial process. There are two reasons for this: First, the data available is not informative enough to infer flux evolution at such fine time scales and, second, we are primarily interested in the average flux over a larger time window.  In contrast, the mole-fraction field, which is directly observed with relatively low measurement error, is only informative of the flux field once the meteorology  (that varies rapidly in space and time) and any chemistry is catered for. To account for time dependence, from now on we add the subscript $t$ to both the mole-fraction field and the interaction function; these become $Y_{2,t}(\cdot)$ and $b_t(\cdot,\cdot)$, respectively.

The flux field is not directly observed; rather, the mole-fraction field is observed at different time points, $t \in \mathcal{T} \equiv \{1,2,\dots,T\}$. Denote the observations at time point $t$ as $\Zvec_{2,t} \equiv (Z_{2,t}(\svec): \svec \in D^O_2)'$, where the finite set $D^O_2 \subset D$. For clarity of exposition we omit the subscript $t$ on $D_2^O$, however our model can readily accommodate time-dependent observation locations. It is reasonable to assume that the observations are conditionally independent and Gaussian, given the true mole-fraction field at the corresponding locations at time $t$. Denote these true mole fractions as $\Yvec_{2,t}^Z \equiv (Y_{2,t}(\svec):\svec \in D^O_2)'$; then the observation model is:
\begin{equation}\label{eq:obs_eq}
(\Zvec_{2,t}\mid Y_{2,t}(\cdot)) \sim \mathcal{N}(\Yvec_{2,t}^Z,\Vmat_t); \quad t \in \mathcal{T},
\end{equation}
where the covariance matrices $\{\Vmat_t:~t \in \mathcal{T}\}$ are diagonal and, in our case, assumed known.

% For each $Y_{2,t}(\cdot)$ we associate an interaction function $b_t(\cdot,\cdot)$ and a discrepancy $\zeta_t(\cdot)$, which we assume contains spatio-temporal correlations. In our case, $b_t(\cdot,\cdot)$ is obtained from a particle dispersion computer model. 

\subsubsection{The mole-fraction field}

The mole-fraction field at time $t$, $Y_{2,t}(\cdot)$, is, to leading order, a linearly transformed version of the flux field $Y_1(\cdot)$, where the linear operator depends on the interaction function. In general, a model for the mole-fraction field follows through conditioning on the flux field:
\begin{equation*}
(\{Y_{2,t}(\cdot):t \in \mathcal{T}\} \mid Y_1(\cdot)) \sim \textrm{Dist}\left(\left\{\int_D b_t(\cdot,\uvec)Y_1(\uvec)\intd \uvec: t \in \mathcal{T}\right\}; \Thetab \right),
\end{equation*}
where $\textrm{Dist}(\{\mu_t(\cdot): t \in \mathcal{T}\}; \Thetab)$ is a possibly non-Gaussian, spatio-temporal (discrete-time) process with mean function $\{\mu_t(\cdot): t \in \mathcal{T}\}$ and higher-order cumulant functions (Section \ref{sec:nonGaussian}) parameterised by $\Thetab$. This conditional distribution is, in reality, extremely complicated, but to date there has been very little effort to adequately characterise it. A natural way forward is to assume \eqref{eq:Y2}, adapted for the presence of the temporal index $t$:
\begin{equation}\label{eq:Y2_time}
Y_{2,t}(\svec) = \int_D b_t(\svec,\uvec)Y_1(\uvec)\intd \uvec + \zeta_t(\svec); \quad \svec \in D, t \in \mathcal{T},
\end{equation}
where $\zeta_t(\cdot),$ for $t \in \mathcal{T}$, has zero expectation. The term $\zeta_t(\cdot)$ in \eqref{eq:Y2_time} is required to capture the spatio-temporal variability in $Y_{2,t}(\cdot)$ that is not explained through the integral operator. Such discrepancy could be due to, for example, incorrect specification of $b_t(\cdot,\cdot)$ in \eqref{eq:Y2_time} or its approximate numerical implementation. 

Since $\{Y_{2,t}(\cdot): t \in \mathcal{T}\},$ is a non-Gaussian spatio-temporal field, modelling and computation are critical considerations. Now, without compromising the non-Gaussian nature of the flux field $Y_1(\cdot)$ (and hence of $Y_{2,t}(\cdot),$ for $t \in \mathcal{T}$), we make the assumption that the additive discrepancy $\zeta_t(\cdot),$ for $t \in \mathcal{T},$ is Gaussian. For this model, the $n$-th order spatial cumulant functions associated with $Y_{2,t}(\cdot),$ for $t \in \mathcal{T}$ and $n \ge 3$ are 
\begin{align*}
\kappa_{Y_{2,t}\dots Y_{2,t}}^n(\svec_1,\dots,\svec_n) &= \int_D\cdots\int_D b_t(\svec_1,\uvec_1)\dots b_t(\svec_n,\uvec_n) \kappa_{Y_1\dots Y_1}^n(\uvec_1,\dots, \uvec_n) \intd \uvec_1\dots\intd \uvec_n; \nonumber \\ &\qquad\qquad\qquad \svec_1,\dots,\svec_n \in D,\label{eq:cumulants2}
\end{align*}
since all cumulants associated with a Gaussian process are zero for orders 3 and higher. 

In practice, each $b_t(\cdot,\cdot)$ is evaluated at a finite set of pairwise discrete locations to yield the matrix $\widetilde\Bmat_t \equiv (b_t(\svec,\uvec): \svec \in D^L_2,\uvec \in D^L_1)$, where $D^L_1$ and $D^L_2$ are finite, possibly different, subsets of $D$ on which we wish to model (and predict) $Y_1(\cdot)$ and $Y_{2,t}(\cdot)$, respectively. Define $\Yvec_1 \equiv (Y_1(\svec): \svec \in D^L_1)', \Yvec_{2,t} \equiv (Y_{2,t}(\svec): \svec \in D^L_2)'$, and $\zetab_t \equiv (\zeta_t(\svec):\svec \in D^L_2)'$. Then \eqref{eq:Y2_time} can be approximated by, $\Yvec_{2,t} \approx \Bmat_t\Yvec_1 + \zetab_t$, where $\Bmat_t \equiv \widetilde\Bmat_t\Deltab$ and $\Deltab$ is a diagonal matrix containing integration weights. Provided $D_2^O \subset D_2^L$, we can obtain $\Yvec_{2,t}^Z$ from $\Yvec_{2,t}$ through a known incidence matrix $\Cmat_t,$ so that $\Yvec_{2,t}^Z = \Cmat_t\Yvec_{2,t}$ for each $t \in \mathcal{T}$. Henceforth, we assume that $D^O_2 \subset D_2^L$.

In the absence of prior insight on the structure of $\zeta_t(\cdot),$ we further assume that its space-time covariance function is separable. Specifically, we assume equally spaced time points, and we let $\bSigma_{\zeta;\tau_2,a,d} \equiv (\cov(\zeta_t(\svec_1),\zeta_{t'}(\svec_2) \mid a,d): t,t' \in \mathcal{T}, \svec_1, \svec_2 \in D^L_2) = \frac{1}{\tau_2}(\Qmat_{\zeta;a}^{t})^{-1} \otimes \Rmat^s_{\zeta;d}$, where $\tau_2$ is a precision parameter; $\Qmat^t_{\zeta;a}$ is the precision matrix corresponding to a first-order auto-regressive discrete-time process \citep[][Chapter 1]{Rue_2005} with auto-regressive parameter $a$; and $\Rmat^s_{\zeta;d} \equiv (\rho_{\zeta}(\|\svec_1 - \svec_2\| \mid d): \svec_1, \svec_2 \in D^L_2)$, where $\rho_{\zeta}(u \mid d) \equiv \exp(-u/d)$, for $u\ge 0$, is the exponential correlation function. The motivation for this choice is primarily computational: In atmospheric trace-gas inversion with ground station data, $|\mathcal{T}|$ is large (several hundreds) while $|D^L_2|$ is typically small and frequently equal to $|D^O_2|$, which is generally less than 10. Hence, $\bSigma_{\zeta;\tau_2,a,d}^{-1}$ will be sparse, although $\Rmat^s_{\zeta;d}$ (and $(\Rmat^{s}_{\zeta;d})^{-1}$) is dense. Other approaches, such as covariance tapering \citep{Kaufman_2008} or dimensionality reduction \citep{Cressie_2008}, might instead be employed to facilitate the computation. 

Note that $\zeta_t(\cdot)$ induces an auto-regressive structure on $Y_{2,t}(\cdot)$. To see this, one can use the alternative representation \citep{Storvik_2002}, $\zeta_t(\cdot) = a\zeta_{t-1}(\cdot) + e_t(\cdot)$, to re-write \eqref{eq:Y2_time} as\ifdetails \footnote{From Example 2 in \citet{Storvik_2002}, if
$$ C_\zeta(r,\hvec) = C_\zeta^r(r)C_\zeta^s(\hvec) $$
and $C_1(r) = a^{|t - t'|}$, then
$$ \zeta_t(\svec) = a \zeta_{t-1}(\svec) + e_t(\svec) $$
where $\textrm{cov}(e_t(\svec)) \propto C_\zeta^s(\hvec)$. Therefore
\begin{align*}
 Y_2(\svec) &= \int_Db_t(\svec,\uvec)Y_1(\uvec) \intd \uvec + a\zeta_{t-1}(\svec) + e_t(\svec) \\
&= \int_D b_t(\svec,\uvec)Y_1(\uvec) + a\left[Y_{2,t-1}(\svec) - \int_D b_{t-1}(\svec,\uvec)Y_1(\uvec)\intd \uvec\right] + e_t(\svec),
\end{align*}
from which the result follows.} \fi
\begin{equation}\label{eq:ARequiv}
Y_{2,t}(\svec) = \int_D (b_t(\svec,\uvec) - a b_{t-1}(\svec,\uvec)) Y_1(\uvec) \intd \uvec + aY_{2,t-1}(\svec) + e_t(\svec); \quad \svec \in D, ~~t \in \mathcal{T},
\end{equation}
where $e_t(\cdot)$ is temporally uncorrelated and has spatial correlation function $\rho_{\zeta}(u\mid d)$. In what follows, we collect the spatial and temporal correlation parameters into the parameter vector $\thetab_2 \equiv (a,d)'$.

\subsubsection{The flux-field model} \label{sec:flux}

The focus now turns to flexible non-Gaussian modelling of $Y_1(\cdot)$, for which we employ a trans-Gaussian-process approach.
%One might make use of the generalised lognormal processes first discussed in \citet{Neyman_1960}, that allow for closed-form expressions of the expectations and covariances \citep{Shimizu_1987}. These processes are constructed by transforming a Gaussian process, $\widetilde Y_1(\svec)$, to $f(Y_1(\svec))$ through a function $f(\cdot)$ that satisfies $f''(x) = A + Bf(x), x\in\mathbb{R}$. It is easy to show that the general solution of this second-order differential equation is a sum of exponentials $(B > 0)$, a sum of trigonometric functions $(B<0)$ or a quadratic expression in $x$ ($B = 0$). This is limiting, although arguably closed-form availability of the expectation and covariance might aid in sampling when carrying out inference using Markov chain Monte Carlo (MCMC). 
For several trace gases, such as methane (considered in Section \ref{sec:semi-empirical}), it is reasonable to assume that the flux is only positive. To this end, we apply a monotonic function that has as domain the positive real line, such that $\widetilde{Y}_1(\cdot) \equiv g_\lambda(Y_1(\cdot))$ is approximately a Gaussian process with mean $\widetilde\mu_1(\cdot) \equiv \xvec(\cdot)'\betab$, and covariance function $\widetilde{C}_{Y_1Y_1}(\cdot,\cdot \mid \thetab_1)$, where $\xvec(\cdot)$ is a vector of covariate functions, $\betab \in \mathbb{R}^p$, and $\lambda$ is a parameter of the transformation $g_\lambda(\cdot)$. 

A leading case of a trans-Gaussian process is the Box--Cox spatial process. Specifically, in line with \cite{deOliveira_1997}, we consider the \emph{power-normal} family of spatial processes for the flux field, obtained through application of the Box--Cox transformation
\begin{align*}
g_\lambda(y) = \left\{\begin{array}{cc} \frac{y^{\lambda}-1}{\lambda}; & \lambda \ne 0   \\
                                         \ln y;& \lambda = 0 \end{array}\right..
\end{align*}
Define the correlation function $\widetilde{R}_{Y_1Y_1}(\cdot,\cdot\mid\thetab_1) \equiv {\tau_1}\widetilde C_{Y_1Y_1}(\cdot,\cdot \mid \thetab_1)$, where $\tau_1$ is a precision parameter. Further, define $\widetilde{\Rvec}_{Y_1Y_1;\thetab_1} \equiv (\widetilde{R}_{Y_1Y_1}(\uvec_1,\uvec_2\mid \thetab_1): \uvec_1,\uvec_2 \in D^L_1)$, $\Xvec \equiv (\xvec(\uvec_1): \uvec_1 \in D^L_1)'$, and $\Yvec_1 \equiv ( Y_1(\uvec_1): \uvec_1 \in D^L_1)'$.  In the case of the Box--Cox family of transformations, the positive-only domain of $g_\lambda(\cdot)$ also restricts the domain of $g_\lambda^{-1}(\cdot)$. In particular, since $g_\lambda^{-1}(\widetilde{y}) = (\lambda\widetilde{y} + 1)^{1/\lambda}$, then $\widetilde{y} > -1/\lambda$ if $\lambda > 0$ and $\widetilde{y} < -1/\lambda$ if $\lambda < 0$. Therefore, the process $\widetilde{Y}_1(\cdot)$ that we specify in this case is a \emph{truncated} Gaussian process. Then the probability density function of $\widetilde{\Yvec}_1 \equiv g_\lambda(\Yvec_1)$ is
\begin{equation*}
p(\widetilde{\Yvec}_1 \mid \betab, \tau_1,\thetab_1,\lambda) = \frac{\left(\frac{\tau_1}{2\pi}\right)^{\frac{|D^L_1|}{2}}|\widetilde{\Rmat}_{Y_1Y_1;\thetab_1}|^{-\frac{1}{2}}}{K_{1;\betab,\tau_1,\thetab_1,\lambda}}\exp{\left(-\frac{\tau_1}{2}(\widetilde{\Yvec}_1 - \Xvec\betab)'\widetilde{\Rvec}_{Y_1Y_1;\thetab_1}^{-1}(\widetilde\Yvec_1 - \Xvec\betab) \right)}T_\lambda(\widetilde\Yvec_1),
\end{equation*} 
where the normalising factor $K_{1;\betab,\tau_1,\thetab_1,\lambda} \le 1$ is a function of all unknown parameters; it is understood that $g_\lambda(\Yvec_1)$ denotes a vector obtained by element-wise application of the Box--Cox transformation; and
\begin{equation}\label{eq:T}
T_\lambda(\widetilde\Yvec_1) \equiv \left\{\begin{array}{cc} \mathbbm{1}(\widetilde\Yvec_1 > -1/\lambda); & \lambda > 0 \\ \mathbbm{1}(\widetilde\Yvec_1 < -1/\lambda); & \lambda < 0 \end{array}. \right.
\end{equation}
In \eqref{eq:T}, $\mathbbm{1}(\cdot)$ is the indicator function as applied to vectors, that is, it returns a one if all the elements of the vector satisfy the condition and a zero otherwise.

From $p(\widetilde{\Yvec}_1 \mid \betab,\tau_1,\thetab_1,\lambda),$ we obtain $p(\Yvec_1 \mid \betab,\tau_1,\thetab_1,\lambda)$ through the transformation $g_\lambda(\cdot)$ to yield the flux-process layer in our hierarchical model,
\begin{align}
p(\Yvec_1 \mid \betab, \tau_1,\thetab_1,\lambda) &=\frac{\left(\frac{\tau_1}{2\pi}\right)^{\frac{|D^L_1|}{2}}|\widetilde{\Rmat}_{Y_1Y_1;\thetab_1}|^{-\frac{1}{2}}}{K_{1;\betab,\tau_1,\thetab_1,\lambda}}\exp{\left(-\frac{\tau_1}{2}(g_\lambda(\Yvec_1) - \Xvec\betab)'\widetilde{\Rvec}_{Y_1Y_1;\thetab_1}^{-1}(g_\lambda(\Yvec_1) - \Xvec\betab) \right)} \nonumber\\
&~~~ \times J_\lambda\mathbbm{1}(\Yvec_1>0),\label{eq:Y1cond}
\end{align}
\noindent where the Jacobian $J_\lambda \equiv \prod_{i=1}^{|D^L_1|} |Y_{1,i}^{\lambda-1}|$. Note that the support of $p(\Yvec_1 \mid \betab,\tau_1,\thetab_1,\lambda)$ is positive, as required. We term \eqref{eq:Y1cond} a Box--Cox spatial model and $Y_1(\cdot)$ a Box--Cox spatial process.

\subsubsection{The parameter model}

Finally, we need to specify prior distributions over the parameters. This should be done with care since, first, the range of the transformed observations changes with the transformation parameter and, second, posterior-distribution impropriety can easily follow from inadequate specification of non-informative prior distributions \citep{Berger_2001}.  In their original paper, \citet{Box_1964} tackled the first issue by considering the conditional prior distribution, $p(\betab,\tau_1,\thetab_1 \mid \lambda) \propto h(\lambda)p(\thetab_1)/\tau_1$, and deducing what $h(\lambda)$ should be for the prior probability measure to be approximately independent of $\lambda$. Their choice, $h(\lambda) = J_\lambda^{p/|D_1^L|}$, is not appropriate in our context where $\Yvec_1$ is not observed. We therefore adopt the alternative of \citet{Pericchi_1981}, who noted that if the conditional prior distribution, $p(\betab,\tau_1,\thetab_1 \mid \lambda) \propto h(\lambda)p(\thetab_1)\tau_1^{\frac{p}{2}-1}$, then $h(\lambda)$ must be constant. 

The prior distribution we choose for the flux-process layer is therefore,
%In order to circumvent these problems, \citet{deOliveira_1997}, based on linearity arguments first put forward in the original paper of \citet{Box_1964}, uses the data-dependent prior distribution \red{Here we have a proble, we cannot make the prior data-dependent as we get a loop in the graphical model}
\begin{equation*}
% p(\betab,\tau_1,\thetab_1,\lambda) \propto \frac{p(\thetab_1)p(\lambda)}{\tau_1 J_\lambda^{p/n}},
p(\betab,\tau_1,\thetab_1,\lambda) \propto p(\thetab_1)p(\lambda)\tau_1^{\frac{p}{2}-1},
\end{equation*}
where $p(\thetab_1)$ and $p(\lambda)$ are marginal prior distributions and, in order to ensure propriety of the posterior distribution, the parameter spaces of $\thetab_1$ and $\lambda$ are bounded \citep{deOliveira_1997}.\ifdetails
\footnote{
Fix a reference value of $\lambda$, say $\lambda_1$ and assume that $z_\lambda \equiv g_\lambda(y)$ is approximately related to $z_{\lambda_1}$ over the range of observations
$$
z_\lambda = const. + l_\lambda z_{\lambda_1}.
$$
Let $\mu_\lambda \equiv \E(z_\lambda)$. Then it follows that $\mu_\lambda = const. + l_\lambda \mu_{\lambda_1}$ and hence
$$
\intd \mu_\lambda = l_\lambda \intd \mu_{\lambda_1}.
$$
This increment dependence is not evident with the parameter $\ln \tau$. Let $\tau_\lambda = \textrm{prec}(z_\lambda)$. Then $\tau_\lambda = l_\lambda^{-2}\tau_{\lambda_1}$ and therefore 
$$
\ln \tau_{\lambda} = const. + \ln \tau_{\lambda_1},
$$
and these increments are hence, approximately, $\lambda$-independent. 

Therefore, in order for the increments associated with the prior density, $h(\lambda)\intd \betab\intd(\ln \tau)$, to be invariant to $\lambda$, one should then set $h(\lambda) = 1/l_\lambda^p$, where $p$ is the number of covariates. \citet{Box_1964} suggested that $l_\lambda = J_\lambda^{1/n}$.

Now consider the Pericchi alternative. The increments associated with the prior of \citet{Pericchi_1981} are 
$$
h(\lambda)\intd\beta\intd(\tau^{p/2}).
$$
Then, using the same linear transformation, we have that $\tau_\lambda = l_\lambda^{-2}\tau_{\lambda_1}$ and hence
$$
\tau_\lambda^{p/2} = l_\lambda^{-p}\tau^{p/2}.
$$
This scaling cancels that due to $\intd \betab$ and hence once can deduce that $h(\lambda)\propto 1$. Note that an increment $\intd (\tau^{p/2})$ is associated with a density $p(\tau) \propto \tau^{\frac{p}{2} - 1}$.

}\fi~For the remaining parameters appearing in the mole-fraction-process layer ($\tau_2,a$ and $d$), we use independent, proper, prior distributions, $p(\tau_2),p(a),$ and $p(d)$, respectively. 

For convenience, we specify bounded uniform distributions for $\thetab_1, \lambda, \ln \tau_2^{-1}, a$, and $\ln d$; see Section \ref{sec:summary} for details.

\subsection{Incorporation of the inventory}\label{sec:inventory}

In atmospheric trace-gas inversion, one is usually supplied with a flux `inventory,' which can be used to reduce uncertainty on inferences on the flux field, $Y_1(\cdot)$. We denote this inventory, available for spatial locations in $D^L_1$, as $\Wvec_1$.  Incorporation of $\Wvec_1$ in the model is akin to the problem of \emph{data assimilation}, where observational data is fused with computer-model output in order to obtain an estimate that is optimal in some sense \citep[e.g.,][]{Wikle_2007}. Standard data assimilation typically assumes that $\E(\Yvec_1) = \Wvec_1$. However, there are two concerns with data assimilation in this context. First, %trace-gas inversion is an ill-posed problem \citep{Michalak_2004}, it is likely that the posterior expectation is highly influenced by the prior expectation, and 
no guarantees are provided on the quality of $\Wvec_1$, which has often been shown to be inaccurate \citep[][amongst others]{Lunt_2015,Miller_2013}. Second, it is unclear how to construct the covariance matrix $\var(\Yvec_1)$; typically a diagonal structure is assumed with elements based on expert judgement \citep[e.g.,][]{Ganesan_2014}, however this is likely to be overly simplistic. Another approach to assimilation is that of \citet{Fuentes_2005}, in which $\Wvec_1$ is treated as a (possibly both additively and multiplicatively biased) observation of $\Yvec_1$. However, in their approach one also has a $\Zvec_1$ that provides a direct (unbiased) observation of $\Yvec_1$. In our context, where direct observations of $\Yvec_1$ are not available, data-driven inferences on the bias terms would be ill-constrained.

We adopt a different approach to the two outlined above, by assuming that $\Wvec_1$ is only informative on the spatial properties of the process. This choice is motivated by the fact that inventories are typically constructed bottom-up from spatially referenced datasets on (say) agricultural productivity and transport infrastructure. The emission factors used in this construction are typically not well constrained, however the main spatial features are still adequately represented. We implement the approach by assuming that $\Wvec_1$ is a noiseless observation of an independent realisation of $(\Yvec_1 \mid \betab,\tau_1, \thetab_1,\lambda)$. Under this modelling assumption, 
$$
p(\Yvec_1 \mid \Wvec_1,\Zvec_2,\betab,\tau_1,\thetab_1,\lambda) \equiv p(\Yvec_1 \mid \Zvec_2,\betab,\tau_1,\thetab_1,\lambda),
$$
due to the conditional independence of $\Yvec_1$ and $\Wvec_1$ when conditioned on $(\betab,\tau_1, \thetab_1,\lambda)$. Hence, this modelling assumption allows us to use Bayesian learning to glean information on $(\betab,\tau_1, \thetab_1,\lambda)$ from $\Wvec_1$. At the same time, inferences on $\Yvec_1$, when conditioned on $\betab,\tau_1,\thetab_1$, and $\lambda$, are purely a function of $\Zvec_2$ and are hence data-driven. We propose that the term \emph{data-feature assimilation} be used to describe this methodology, to distinguish it from standard data assimilation.

In this application, we envision that the transformation parameter $\lambda$, and consequently all third-order and higher-order joint cumulants of $\Yvec_1$, will be highly sensitive to $\Wvec_1$. Hence, this procedure is analogous to the use of a \emph{training image} to estimate third-order and higher-order cumulants prior to carrying out spatial prediction \citep{Dimitrakopoulos_2010}.

As a result of the modelled conditional independence between $\Yvec_1$ and $\Wvec_1$, we augment the conditional distribution \eqref{eq:Y1cond} as follows:
\begin{align}
p(\Yvec_1,\Wvec_1 \mid \betab, \tau_1,\thetab_1,\lambda) &= p(\Yvec_1 \mid \betab, \tau_1,\thetab_1,\lambda)p(\Wvec_1 \mid \betab, \tau_1,\thetab_1,\lambda) \nonumber \\
&= \frac{\left(\frac{\tau_1}{2\pi}\right)^{|D^L_1|}|\widetilde{\Rmat}_{Y_1Y_1;\thetab_1}|^{-1}}{K_{2;\betab,\tau_1,\thetab_1,\lambda}}\exp{\left(-\frac{\tau_1}{2}\sum_{j=1}^2(\Gvec_{\lambda,j} - \Xvec\betab)'\widetilde{\Rvec}_{Y_1Y_1;\thetab_1}^{-1}(\Gvec_{\lambda,j} - \Xvec\betab) \right)}\nonumber \\
&~~~ \times \left(\prod_{j=1}^2J_{\lambda,j}\right)\mathbbm{1}(\Yvec_1 > 0)\mathbbm{1}(\Wvec_1 > 0), \label{eq:Y1_cond}
\end{align}
where ${K_{2;\betab,\tau_1,\thetab_1,\lambda}}$ is a normalising factor, $\Gvec_{\lambda,1} \equiv g_\lambda(\Yvec_1), \Gvec_{\lambda,2} \equiv g_\lambda(\Wvec_1)$, $J_{\lambda,1} \equiv J_\lambda$ in \eqref{eq:Y1cond}, and $J_{\lambda,2} \equiv \prod_{i=1}^{|D^L_1|} |W_{1,i}^{\lambda-1}|$. For notational convenience, we frequently consider the joint vector $\underline\Yvec_{1} \equiv (\Yvec_1',\Wvec_1')'$; then the conditional distribution \eqref{eq:Y1_cond} becomes
\begin{align}
p(\underline\Yvec_{1} \mid \betab, \tau_1,\thetab_1,\lambda) &= \frac{\left(\frac{\tau_1}{2\pi}\right)^{|D^L_1|}|\underline{\widetilde{\Rmat}}_{Y_1Y_1;\thetab_1}|^{-1/2}}{K_{2;\betab,\tau_1,\thetab_1,\lambda}}\exp{\left(-\frac{\tau_1}{2}(\underline\Gvec_{\lambda} - \underline\Xvec\betab)'\underline{\widetilde{\Rvec}}_{Y_1Y_1;\thetab_1}^{-1}(\underline\Gvec_{\lambda} - \underline\Xvec\betab) \right)} \nonumber\\
&~~\times \underline J_{\lambda}\mathbbm{1}(\underline\Yvec_1 > 0).\label{eq:Y1_cond2}
\end{align}
In \eqref{eq:Y1_cond2}, $\underline\Gvec_{\lambda} \equiv (\Gvec_{\lambda,1}',\Gvec_{\lambda,2}')'$, $\underline\Xvec \equiv (\Xvec',\Xvec')'$, $\underline J_{\lambda} \equiv \prod_{i=1}^{2|D_L|}|\underline{Y}_{1,i}^{\lambda-1}|$, and $\underline{\widetilde\Rvec}_{Y_1Y_1;\thetab_1} \equiv \textrm{bdiag}(\widetilde\Rvec_{Y_1Y_1;\thetab_1},\widetilde\Rvec_{Y_1Y_1;\thetab_1})$, where $\textrm{bdiag}(\cdot)$ creates a block-diagonal matrix from its arguments.

\subsection{Summary of the hierarchical model}\label{sec:summary}

\begin{figure}[t!]
\begin{center}

\begin{tikzpicture}[scale=0.8,every node/.style={transform shape}]
%[inner sep=1mm] 
[line width=1pt]

\path (-1,2) node (z21) [shape=circle,minimum size=1cm,draw] {\small $\Zvec_{2,1}$};
\path (1,2) node (z22) [shape=circle,minimum size=1cm,draw] {\small $\Zvec_{2,2}$};
\path (6,2) node (z2T) [shape=circle,minimum size=1cm,draw] {\small $\Zvec_{2,T}$};
\path (3.5,2) node (zdots) [shape=rectangle,minimum size=1cm,draw=none] {\small $\dots\dots$};

\path (-1,0) node (y21) [shape=circle,minimum size=1cm,draw] {\small $\Yvec_{2,1}$};
\path (1,0) node (y22) [shape=circle,minimum size=1cm,draw] {\small $\Yvec_{2,2}$};
\path (6,0) node (y2T) [shape=circle,minimum size=1cm,draw] {\small $\Yvec_{2,T}$};
\path (3.5,0) node (ydots) [shape=rectangle,minimum size=1cm,draw=none,minimum width=2cm] {\small $\dots\dots$};

\path (4,-2) node (y1) [shape=circle,minimum size=1cm,draw] {\small $\Yvec_{1}$};
%\path (-3,-3) node (a) [shape=circle,minimum size=1cm,draw] {\small $a$};
%\path (-1.5,-3) node (d) [shape=circle,minimum size=1cm,draw] {\small $d$};
\path (-2,-3) node (theta2) [shape=circle,minimum size=1cm,draw] {\small $\thetab_2$};
\path (0,-3) node (sigma2) [shape=circle,minimum size=1cm,draw] {\small $\tau_2$};

\path (6,-2) node (y1inv) [shape=circle,minimum size=1cm,draw] {\small $\Wvec_1$};
\path (2,-4) node (tau) [shape=circle,minimum size=1cm,draw] {\small $\tau_1$};
\path (4,-4) node (beta) [shape=circle,minimum size=1cm,draw] {\small $\betab$};
\path (6,-4) node (theta1) [shape=circle,minimum size=1cm,draw] {\small $\thetab_1$};

\path (8,-4) node (lambda) [shape=circle,minimum size=1cm,draw] {\small $\lambda$};

%\node (zetapars) at (-1.5,-3) [draw,thick,minimum width=5cm,minimum height=2cm] {};

 \draw [->] (y21) to (z21);
 \draw [->] (y22) to (z22);
 \draw [->] (y2T) to (z2T);
 \draw [->] (y1) to (y21);
 \draw [->] (y1) to (y22);
 \draw [->] (y1) to (y2T);

 \draw [->] (y21) to (y22);
 \draw [->] (y22) to (ydots);
 \draw [->] (ydots) to (y2T);

 \draw [->] (tau) to (y1);
 \draw [->] (tau) to (y1inv);
 \draw [->] (beta) to (y1);
 \draw [->] (beta) to (y1inv);

 \draw [->] (theta1) to (y1);
 \draw [->] (theta1) to (y1inv);

 \draw [->] (lambda) to (y1);
 \draw [->] (lambda) to (y1inv);

%\draw [->] (zetapars) to (y21);
%\draw [->] (zetapars) to (y22);
%\draw [->] (zetapars) to (y2T);

% \draw [->] (a) to (y2T);
% \draw [->] (a) to (y21);
% \draw [->] (a) to (y22);

% \draw [->] (d) to (y2T);
% \draw [->] (d) to (y21);
% \draw [->] (d) to (y22);

\draw [->] (theta2) to (y2T);
\draw [->] (theta2) to (y21);
\draw [->] (theta2) to (y22);

\draw [->] (sigma2) to (y2T);
\draw [->] (sigma2) to (y21);
\draw [->] (sigma2) to (y22);

% \draw [->] () to (y);
\end{tikzpicture}

\end{center}
\caption{Graphical representation of the hierarchical model, where $\{\Zvec_{2,t}: t \in \mathcal{T}\}$ are the mole-fraction observations, $\{\Yvec_{2,t}: t \in \mathcal{T}\}$ are the mole fractions, $\Yvec_1$ is the flux field, $\Wvec_1$ is the flux-field inventory, $\thetab_2$ are mole-fraction-discrepancy spatio-temporal correlation parameters, $\tau_2$ is the mole-fraction-discrepancy precision parameter, $\tau_1$ is the flux precision parameter, $\betab$ are the flux-field regression coefficients, $\thetab_1$ are the flux-field correlation parameters, and $\lambda$ is the Box--Cox transformation parameter.}\label{fig:graphical_model}
\end{figure}
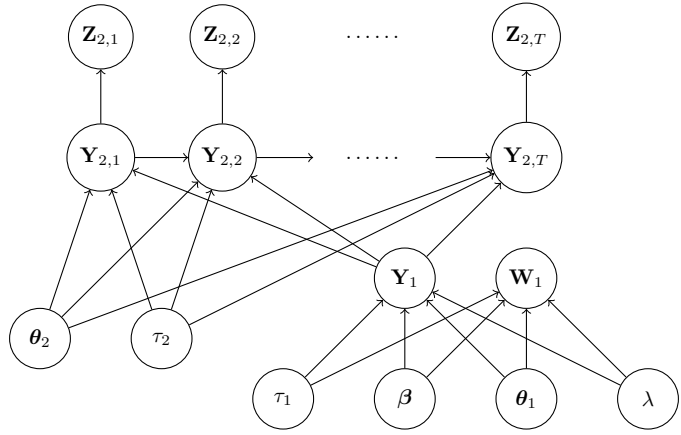

The graphical model that we construct is given in Fig.~\ref{fig:graphical_model}, where for simplicity we omit the discrepancies $\{\zetab_t: t\in \mathcal{T}\}$ and instead show the auto-regressive structure on $\{\Yvec_{2,t}: t \in \mathcal{T}\}$ that can be obtained from \eqref{eq:ARequiv}. In this graphical model we illustrate the time evolution of the mole-fraction field, however it is convenient, both for notational purposes and for inference in Section \ref{sec:inference}, to construct vectors and matrices that are blocked with time. 

Define $\Zvec_2 \equiv (\Zvec_{2,t}': t \in \mathcal{T})'$, $\Cmat \equiv \textrm{bdiag}(\{\Cmat_{t}: t \in \mathcal{T}\})$, $\Vmat \equiv \textrm{bdiag}(\{\Vmat_t : t \in \mathcal{T}\})$, $\Yvec_2 \equiv (\Yvec_{2,t}': t \in \mathcal{T})'$, and $\Bmat \equiv (\Bmat_{t}': t \in \mathcal{T})'$. Then the hierarchical model is:
\begin{align*}
  \textrm{Observation model (mole fraction):} & \qquad (\Zvec_{2} \mid \Yvec_{2}) \sim \mathcal{N}(\Cmat \Yvec_2, \Vmat), \\
\textrm{Process model 2 (mole fraction):}     & \qquad (\Yvec_2 \mid \Yvec_1,\thetab_2,\tau_2) \sim \mathcal{N}(\Bmat\Yvec_1,\Sigmamat_{\zeta;\tau_2,a,d}),  \\
\textrm{Process model 1 (flux):}     & \qquad (\Yvec_1 \mid \betab,\tau_1,\thetab_1,\lambda) \sim \mathcal{BC}\left(\Xvec\betab,\frac{1}{\tau_1}\widetilde\Rmat_{Y_1Y_1;\thetab_1},\lambda\right), \\
\textrm{Inventory model (flux):}     & \qquad (\Wvec_1 \mid \betab,\tau_1,\thetab_1,\lambda) \sim \mathcal{BC}\left(\Xvec\betab,\frac{1}{\tau_1}\widetilde\Rmat_{Y_1Y_1;\thetab_1},\lambda\right), \\
& \qquad \textrm{independent of Process model 1,}\\
\textrm{Parameter model 2 (mole-fraction):}     & \qquad \ln(\tau_2^{-1}) \sim \mathcal{U}(\gamma_{\tau_2^{-1}}^l,\gamma_{\tau_2^{-1}}^u), \qquad a \sim \mathcal{U}(\gamma_a^l,\gamma_a^u), \\ & \qquad \ln{d} \sim \mathcal{U}(\gamma_d^l, \gamma_{d}^u),\\
\textrm{Parameter model 1 (flux):}  & \qquad p(\tau_1) \propto \tau_1^{\frac{p}{2}-1}, \qquad p(\betab) \propto 1, \qquad \lambda \sim \mathcal{U}(\gamma_\lambda^l,\gamma_\lambda^u), \\ & \qquad \theta_{1i} \sim \mathcal{U}(\gamma_{\theta_{1i}}^l,\gamma_{\theta_{1i}}^u); \quad i = 1,2,\dots,n_{\thetab_1},
\end{align*}
where $\mathcal{BC}$ is an abbreviation for `Box--Cox': If $\Dvec \sim \mathcal{BC}(\muvec,\Sigmamat,\lambda)$, then the probability density function of $g_\lambda(\Dvec)$ is proportional to the multivariate normal density function with mean $\muvec$ and covariance matrix $\Sigmamat$, multiplied by $T_\lambda(\Dvec)$ given by \eqref{eq:T}. The distribution $\mathcal{U}(\gamma_\eta^l,\gamma_\eta^u)$ is the bounded uniform distribution over $\eta$ with limits $\gamma_\eta^l$ and $\gamma_\eta^u$, and $n_{\thetab_1}$ is the number of elements in the vector $\thetab_1$. For trace-gas inversion of methane, all hyper-parameters appearing in the parameter models can be elicited from physical considerations \citep[e.g.,][]{Ganesan_2015}.

%In Section \ref{sec:nonGaussian} we showed the flexibility in expressing non-Gaussianity in $Y_2(\cdot)$ through  when constructing a model using a conditional approach. 

\section{Inference}\label{sec:inference}

Since the normalising constant of the truncated multivariate normal density function is intractable, inference using the model in Section \ref{sec:summary} is problematic. Such models, where the normalising constants of both the likelihood and the posterior distribution are intractable, have been called \emph{doubly intractable} \citep{Murray_2006}. The approach of \citet{deOliveira_1997} is to implicitly assume that the truncated volume is approximately 0. Then, in \eqref{eq:Y1cond} and \eqref{eq:Y1_cond2}, the factor $K_{i;\betab,\tau_1,\thetab_1,\lambda} \approx 1,$ for $i = 1,2$, independently of $\{\betab,\tau_1,\thetab_1,\lambda\}$. From here on we assume that $K_{i;\betab,\tau_1,\thetab_1,\lambda} = 1,$ for $i = 1,2$.

The Box--Cox transformation can only induce normality in $\widetilde{\Yvec}_1$ if $\lambda = 0$ \citep{Poirer_1978}. In the  univariate case, truncation is kept to a minimum when the mean of $\widetilde Y_1$ is large and the coefficient of variation of $\widetilde Y_1$ is small \citep{Freeman_2006}. \citet{Draper_1969} conclude that even though the Box--Cox transformation does not achieve exact Gaussian multivariate normality, it has a beneficial effect on third-order cumulants (i.e., skewness); we believe that %showed that when the model is misspecified (i.e., the normalising factor is approximated to one), both estimates or inferences over the transformation parameter $\lambda$ are such that the skewness in the distribution of the transformed parameter is minimised. Hence, the Box--Cox transformation with the normality assumption helps to regularise the data even under misspecification; 
this is the main motivation for its continued use in applications, such as in this article. %However, caution should be exercised by assessing the degree of truncation and hence of mis-specification.

Once we assert that the truncation effect in \eqref{eq:Y1_cond2} is small, then the inferential approach can be based on the standard Gibbs sampler. We can then also analytically marginalise out the variables $\tau_1$, $\betab$, and $\Yvec_2$. There are several reasons why marginalisation is useful, especially when using improper prior distributions \citep[see][for a discussion]{Berger_1999}, however our prime motivation here is to improve the convergence properties of the sampler \citep{Dyk_2008}.

The required full conditional distributions are
\begin{align*}
& p(\tau_2,\thetab_2 \mid \Zvec_2, \Yvec_1), \\
& p(\Yvec_1 \mid \Zvec_2,  \Wvec_1, \thetab_2,\tau_2,\thetab_1, \lambda), \\
& p(\thetab_1,\lambda \mid \underline\Yvec_1),
\end{align*}
where in the Gibbs sampler we block sample $\{\tau_2,\thetab_2\}$ and $\{\thetab_1,\lambda\}$. Here we outline some features of these full conditional distributions; further details are given in \ref{sec:CondDist}.

The first full conditional distribution, $p(\tau_2,\thetab_2 \mid \Zvec_2, \Yvec_1) \propto p(\Zvec_2 \mid \tau_2,\thetab_2,\Yvec_1)p(\tau_2)p(\thetab_2)$, which requires the marginalisation,
\begin{equation}\label{eq:Z2}
p(\Zvec_2 \mid \tau_2,\thetab_2,\Yvec_1) = \int p(\Zvec_2 \mid \Yvec_2) p(\Yvec_2 \mid \tau_2,\thetab_2,\Yvec_1)\intd \Yvec_2.
\end{equation}
This is straightforward since both $p(\Zvec_2 \mid \Yvec_2)$ and $p(\Yvec_2 \mid \tau_2,\thetab_2,\Yvec_1)$ are multivariate Gaussian distributions. The resulting conditional distribution of $(\tau_2,\thetab_2)$ requires the Cholesky decomposition of $\bSigma_{\zeta;\tau_2,a,d}^{-1}$ or other sparse variants of it that can easily be computed due to the imposed auto-regressive structure on $\zeta_t(\cdot), t \in \mathcal{T}$. The full conditional density function is given in \eqref{eq:expand_exponent} and we sampled from it using a slice sampler \citep{Neal_2003}.

The second full conditional distribution, 
\begin{equation} \label{eq:Y1_cond3}
p(\Yvec_1 \mid \Zvec_2, \Wvec_1,\thetab_2,\tau_2,\thetab_1,\lambda) \propto p(\Zvec_2 \mid \tau_2,\thetab_2,\Yvec_1)p(\underline\Yvec_{1} \mid \thetab_1,\lambda),
\end{equation}
also requires marginalisation:
\begin{equation*}
p(\underline \Yvec_1 \mid \thetab_1, \lambda) = \int p(\underline\Yvec_{1} \mid \betab,\tau_1,\thetab_1,\lambda)p(\betab,\tau_1)\intd\betab \intd\tau_1,
\end{equation*}
where the distribution of $(\underline\Yvec_{1} \mid \betab,\tau_1,\thetab_1,\lambda)$ is given by \eqref{eq:Y1_cond2}. This is the integrated likelihood function of $\{\lambda, \thetab_1\}$ \citep{Berger_2001}, which partially depends on $\underline\Yvec_{1}$ through the sum of squared residuals,
\begin{equation}\label{eq:S2}
S^2_{\thetab_1,\lambda} \equiv (\underline\Gvec_{\lambda} - \underline\Xvec\hat\betab_{\thetab_1,\lambda})'\underline{\widetilde{\Rmat}}_{Y_1Y_1;\thetab_1}^{-1}(\underline\Gvec_{\lambda} - \underline\Xvec\hat\betab_{\thetab_1,\lambda}).
\end{equation}
In \eqref{eq:S2}, $\hat\betab_{\thetab_1,\lambda}$ is the generalised least squares estimate of $\betab$ conditioned on $\thetab_1$ and $\lambda$; that is, 
\begin{equation}\label{eq:betaGLS}
\hat\betab_{\thetab_1,\lambda} = (\underline\Xvec'\underline{\widetilde{\Rmat}}_{Y_1Y_1;\thetab_1}^{-1}\underline\Xvec)^{-1}\underline\Xvec'\underline{\widetilde\Rmat}_{Y_1Y_1;\thetab_1}^{-1}\underline\Gvec_{\lambda}.
\end{equation}
 The sum of squared residuals, $S^2_{\thetab_1,\lambda}$, needs to be non-zero everywhere for \eqref{eq:Y1_cond3} to be proper (see \eqref{eq:Y1_logcond} in \ref{sec:CondDist}). A sufficient condition for this is that there does not exist a $\betab$ such that $\Wvec_1 - \Xmat\betab = \zerob$. This condition is difficult to violate in practice (unless the inventory is used as a covariate) and, thus, propriety of the conditional distribution is practically guaranteed.

% which is unbounded from this choice of prior distribution \citep[][p.~257, item 9.29]{OHagan_2000}. 
%Propriety of the conditional distribution is assured because of the first term in the product of \eqref{eq:Z2}, which is multivariate Gaussian with full-rank covariance matrix and thus proper \red{Prove this}. 
Since the gradient of this full conditional distribution can be easily found (see \ref{sec:CondDist}), we implemented a Hamiltonian Monte Carlo (HMC) sampler for sampling from this conditional distribution \citep{Neal_2011}. Previous work on this application \citep{Zammit_2015} indicated an important benefit of HMC sampling. The gradient information of the conditional density function of $\Yvec_1$ improved the mixing properties of the MCMC chains when compared to other samplers, such as slice samplers or Metropolis samplers. Since we analyse several models in this work, we auto-adapt the HMC sampler's step size in the burn-in period, in order to achieve suitable acceptance ratios.

The third full conditional distribution we require is $p(\thetab_1,\lambda \mid \underline\Yvec_1)$, which recall is guaranteed to be proper under the chosen prior distribution. This conditional distribution can be obtained by following the approach in \cite{deOliveira_1997}, after modifying it slightly to include $\Wvec_1$ in addition to $\Yvec_1$. In summary, the distribution of $(\betab,\tau_1 \mid \underline\Yvec_1,\thetab_1,\lambda)$ is multivariate Normal-Gamma and thus has a known normalising constant in terms of all the other parameters. Hence, the required conditional distribution can be found, up to a constant of proportionality, by writing it out as
$$ p(\thetab_1,\lambda \mid \underline\Yvec_1) = \frac{p(\betab,\tau_1,\thetab_1,\lambda \mid \underline\Yvec_1)}{p(\betab,\tau_1 | \underline\Yvec_1,\thetab_1,\lambda)}. $$
We sampled from this conditional distribution using a slice sampler \citep{Neal_2003}.

\section{Trace-gas inversion in the UK and Ireland}\label{sec:semi-empirical}

% In this section we apply the Box--Cox inverse framework for methane flux estimation in the UK and Ireland in two settings. First, in Section \ref{sec:semi-empirical} we carry out a controlled experiment where we assume the flux field is known (identical to the inventory) and simulate the observations at four stations from the inventory through the transport model. Second, in Section \ref{sec:real-data} we apply the algorithm to real data obtained from these stations.

In this section we consider the problem of methane flux inversion in the UK and Ireland using the model of Section \ref{sec:summary} and the Gibbs sampler of Section \ref{sec:inference}. In Sections \ref{sec:OSSE} and \ref{sec:MCMC} we describe the overall simulation setup and experimental conditions, while the results are discussed in Section \ref{sec:results}.

% \subsection{Realistic simulation study}\label{sec:semi-empirical}

\subsection{Observation system simulation experiment (OSSE) setup} \label{sec:OSSE}

The main methane emissions inventory we use is based on the UK National Atmospheric and Emissions Inventory \citep[NAEI, UK][]{NAEI} and the Emissions Database for Global Atmospheric Research 4.2 \citep[EDGAR,][]{Edgar}; see \citet{Ganesan_2015} for details. 

\begin{figure}[!t]
\begin{center}
\includegraphics[width=6.0in]{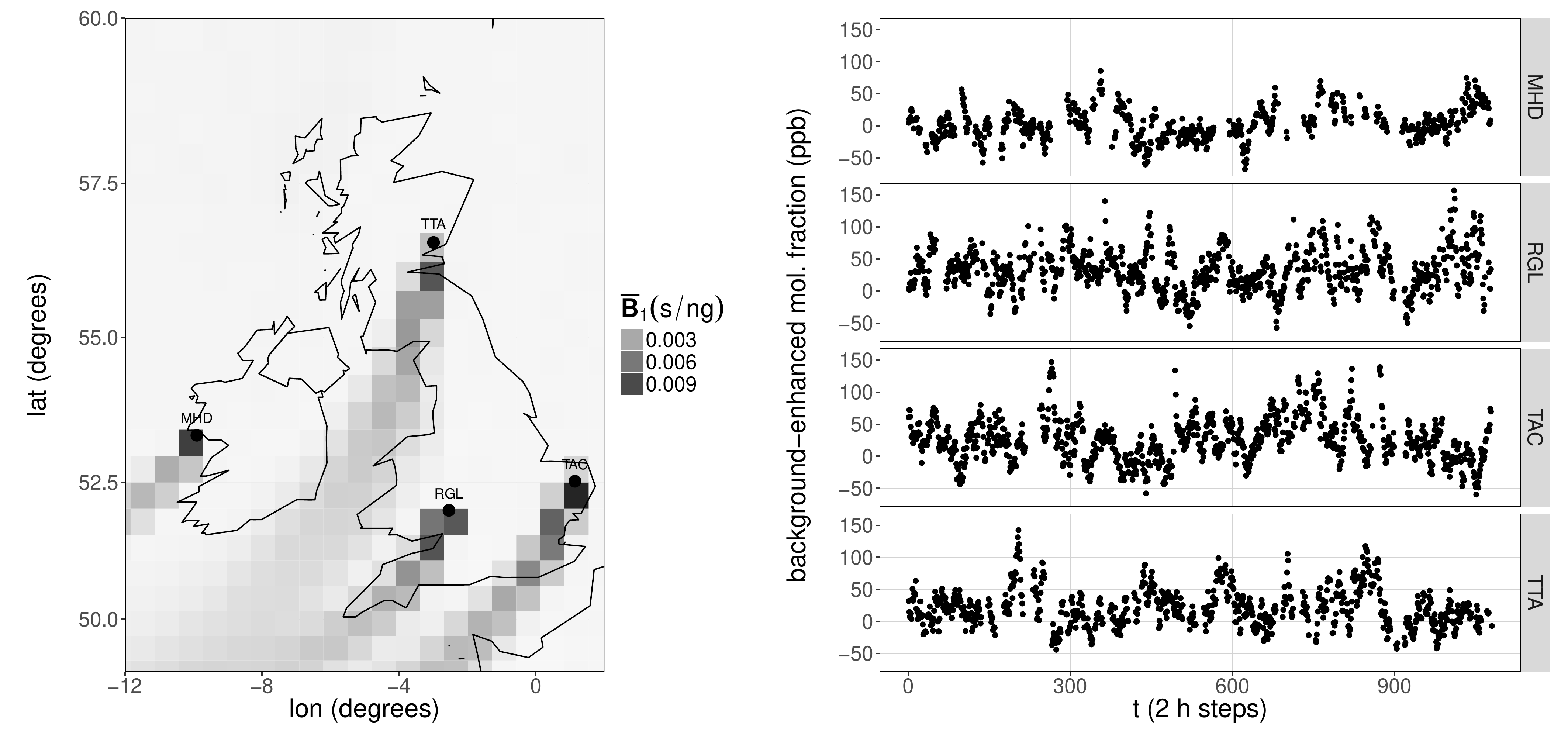}  
	\caption{Left panel: The region of interest (UK and Ireland) and the station locations considered (black dots), that is, Mace Head (MHD), Ridge Hill (RGL), Tacolneston (TAC), and Angus (TTA). The quantity depicted is the vector $\overline \Bmat_1 = \oneb'\Bmat_1$ mapped out geographically; this plot superimposes the sensitivity of mole fraction to the flux at the first time point at the four stations to give a visual summary of $\Bmat_1$. Right panel: The four simulated measurement time series at the four station locations of background-enhanced methane mole fraction. The first time point corresponds to January 01 2014 at 00:00.} \label{fig:sim_setup}
\end{center}
\end{figure}

Since methane is a long-lived gas, we expect that the relationship between the flux and the mole-fraction is linear (Section \ref{sec:inversion_intro}). The matrices used for the linear mapping, $\{\Bmat_t : t \in \mathcal{T}\}$, were constructed from $b_t(\svec,\uvec)$ for $t \in \mathcal{T},$ $\svec \in D^O_2,$ and $\uvec \in D^L_1,$ using the UK Met Office's Lagrangian Particle Dispersion Model (LPDM), NAME. The domain $D^L_1$ was defined as a lon-lat grid, with each grid cell of size 0.7$^\circ$ $\times$ 0.5$^\circ$; here, $|D^L_1| = 122$. The domain $D^L_2$ was chosen to be $D^O_2$, the set of four locations of ground stations recording methane mole fraction in the UK and Ireland; hence $|D^L_2| = 4$. These four stations are located in Mace Head (Ireland), Ridge Hill (England), Tacolneston (England), and Angus (Scotland). The temporal domain $\mathcal{T}$ contains every 2 h interval between January 01 2014 at 00:00 and April 01 2014 at 00:00; a total of 1080 time intervals. Since the Scottish Highlands experience much lower methane emissions than the rest of the UK, we constructed $\Xvec$ with two columns; the first column contains ones in rows corresponding to grid cells above 56.4$^\circ$ latitude (corresponding to the latitudes of the Scottish Highlands) and zeros otherwise, and the second column contains ones in rows corresponding to grid cells below 56.4$^\circ$ and zeros otherwise. The domain of interest $D$, station locations $D_2^O$, and the sum over rows of each column of $\Bmat_1$, are shown in Fig.~\ref{fig:sim_setup}, left panel.

In order to mimic a typical regional inversion study, we carried out what is known as an \emph{observing system simulation experiment} (OSSE), where observations (including measurement error) are forward-simulated using the bivariate model together with a known flux field. Initially, we assumed that the true flux $\Yvec_1$ was equal to the UK and Ireland inventory $\Wvec_1$  (which therefore was not simulated from a Box--Cox spatial model). We further let $\tau_2 = 0.01$ ppb$^{-2}$, $a = 0.9,$ and $d = 2.5^\circ$, in line with what is expected in practice \citep{Zammit_2015}. We next simulated the mole-fraction field using a discrete approximation to \eqref{eq:Y2_time} with the matrices $\{\Bmat_t: t \in \mathcal{T}\}$ supplied from NAME. Finally, the observations were simulated from \eqref{eq:obs_eq}. We assumed that all the matrices in $\{\Vmat_t : t \in \mathcal{T}\}$ were equal to the identity matrix and that data was missing in the OSSE at the same times and locations that data was indeed missing from the stations whose locations we use. The missingness was used to construct the incidence matrices $\{\Cmat_t : t \in \mathcal{T}\}$. 

In reality, the detected methane mole fraction is the sum of the contributions from regional emissions, and a background level (around 1800 ppb) that is not accounted for by the LPDM since it is run for a temporal horizon of only 30 days. This background level varies both in space (e.g., with latitude) and time (e.g., seasonally) and can be included in our model through $\E(\zeta_t(\cdot))$. In this OSSE, we are implicitly assuming that the background is known and that the discrepancy has been adequately corrected for. Therefore, $\E(\zeta_t(\cdot)) = 0$, and henceforth we refer to $Y_{2,t}(\svec)$ as a background-enhanced value of methane mole fraction. Estimation of background mole-fractions is discussed in \citet{Ganesan_2015}. The simulated observations from the four stations are depicted in Fig.~\ref{fig:sim_setup}, right panel. Negative measurements are present in these time series, primarily due to the Gaussian discrepancy $\{\zeta_t(\cdot): t \in \mathcal{T}\}$. In practice, negative residual mole fractions can arise from incorrect characterisation of the background process. Note that the 30-day temporal horizon used in NAME to obtain the required interaction functions is unrelated to the window we use for flux inference (namely, 3 months).

We considered six models for the flux field, namely a Box--Cox spatial process (Model 1), a lognormal spatial process (Box--Cox parameter $\lambda$ fixed to 0; Model 2), a truncated Gaussian process ($\lambda$ fixed to 1 but with positivity imposed; Model 3), and the same three sequences of models but with no spatial correlation assumed (Models 4--6). Specifically, for Models 1--3, we let the transformed flux-field covariance function, $\widetilde C_{Y_1Y_1}(\uvec_1,\uvec_2 | \thetab_1)$, be given by
\begin{equation*}\label{eq:C11b}
\widetilde C_{Y_1Y_1}(\uvec_1,\uvec_2 \mid \thetab_1) = \frac{1}{\tau_1}\exp\left(- \theta_{11} \| \uvec_1 - \uvec_2 \|^{\theta_{12}} \right),
\end{equation*}
where $\thetab_1 \equiv (\theta_{11},\theta_{12})'$, and $\theta_{11} > 0, 0 < \theta_{12} < 2$; for Models 4--6, we let $\widetilde C_{Y_1Y_1}(\uvec_1,\uvec_2 | \thetab_1) = \frac{1}{\tau_1}\mathbbm{1}(\uvec_1=\uvec_2)$. Using prediction performance measures, these models will allow us to weigh the importance of modelling non-Gaussianity against that of spatial modelling. %Such an analysis is of interest since both non-Gaussianity and the use of spatial correlations result in increased difficulties with regards to computation and interpretation. 
Note that Models 1--6 are all stationary models; while this may be a reasonable assumption when modelling the flux at this resolution on a domain of this size, more complex models will be needed for continental (and upwards) domains. In the general case, one might ensure that the employed model is representative of the flux by fitting the flux model to the inventory and looking at simple diagnostics (e.g., through leave-one-out cross-validation).

\begin{figure}[!t]
\begin{center}
\includegraphics[width=6.0in]{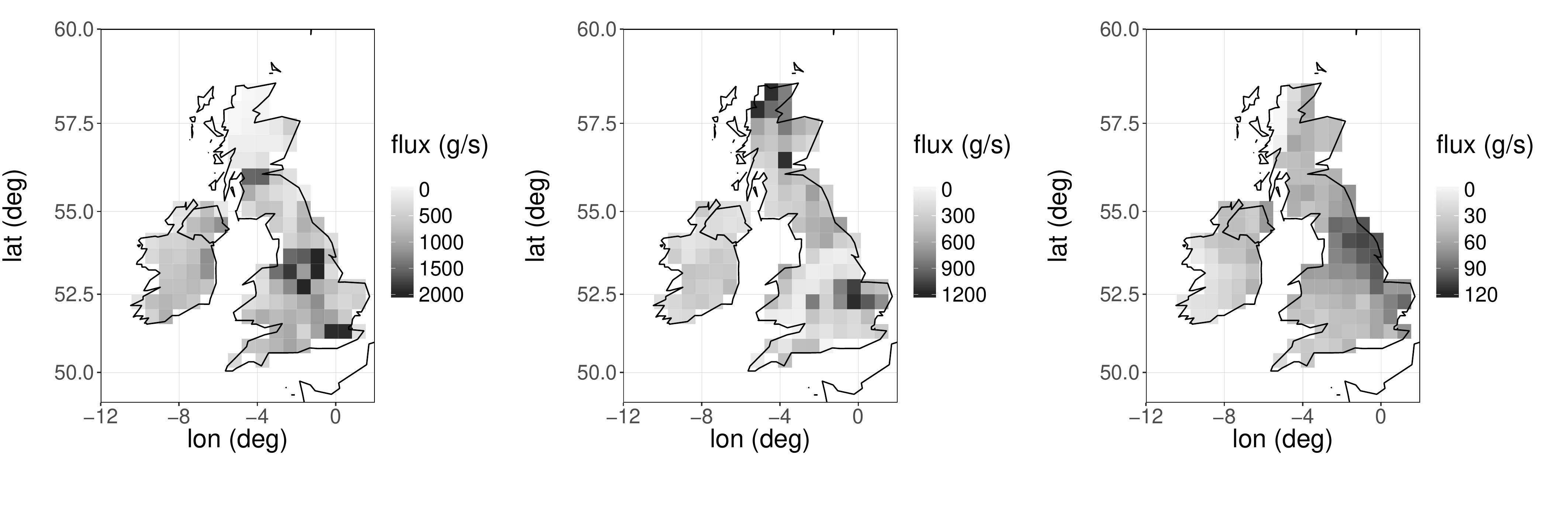}  
	\caption{Inventories used in the study both for OSSE-data generation and data-feature assimilation experiments: The correct inventory of the UK and Ireland (left panel), an incorrect inventory taken from mainland Europe but with similar spatial properties to the UK (centre panel), and an incorrect inventory taken from Northern Australia with highly dissimilar spatial properties (right panel).} \label{fig:all_inventories}
\end{center}
\end{figure}

In order to assess the role the inventory plays in the inference, we re-ran the algorithm with Model 1 using the same observations $\Zvec_2$, but we replaced $\Wvec_1$ with one of two `incorrect' inventories extracted from EDGAR in different world regions; see Fig.~\ref{fig:all_inventories}. The first inventory, $\Wvec_1^*$, was taken from mainland Europe and has second-order properties that are qualitatively similar to those of the UK and Ireland. The second inventory was taken from Northern Australia. Since methane emissions are much lower in this latter region, we shifted and scaled the emissions in this inventory so that they have the same mean and variance as $\Wvec_1$; we denote this inventory as $\Wvec_1^{**}$ . We denote the models utilising these two inventories by Model $1^*$ and Model $1^{**}$, respectively.  

Finally, since it is known that the lognormal spatial process is an adequate model for emissions in the UK and Ireland \citep{Zammit_2015}, we re-ran the OSSE with Models 1--6 assuming that the true emissions field is given by the inventory from Northern Australia (Fig.~\ref{fig:all_inventories}, right panel), while for simplicity keeping the measurement locations and meteorology unchanged. Upon comparing the panels of Fig.~\ref{fig:all_inventories}, it is readily apparent how different the right-hand panel is, and that a lognormal process will not be suitable for modelling the flux field. In this case, the added flexibility of the trans-Gaussian (here, Box--Cox) process is important.

\subsection{MCMC details and diagnostics} \label{sec:MCMC}

%, alleviate concerns that the diagnostics are highly influenced by use of a `perfect' inventory, we used the same observations $\Zvec_2$ but replaced $\Wvec_1$ used in Model 1  with an inventory $\Wvec_1^*$ from mainland Europe; we denote this model as Model $1^{*}$. 

We used bounded uniform prior distributions that encompass physically plausible values on parameter transformations. For the mole-fraction parameters, we used `Parameter model 2' in Section \ref{sec:summary}, and we set $\gamma_{\tau_2^{-1}}^l = -2, \gamma_{\tau_2^{-1}}^u = 20, \gamma_{a}^l = -1, \gamma_{a}^u = 1, \gamma_{d}^l = \ln(0.1)$, and $\gamma_{d}^u = \ln(5)$. For the flux-field parameters, we used `Parameter model 1', and we set $\gamma_{\theta_{11}}^l = 0, \gamma_{\theta_{11}}^u = 2, \gamma_{\theta_{12}}^l = 0, \gamma_{\theta_{12}}^u = 2, \gamma_{\lambda}^l = -3,$ and $\gamma_{\lambda}^u = 3$ as in \citet{deOliveira_1997}.

For each model, we ran 10 parallel MCMC chains with 12000 samples each. We used the first 1000 samples for HMC adaptation, during which the step size used for discretisation of the Hamiltonian dynamics was updated at a decreasing rate of adaptation in order to achieve an acceptance ratio between 30\% and 80\%. These samples, together with a further 7000, were discarded to ensure adequate burn-in. We then took every 10-th sample from the resulting chains, to give a total of 4000 posterior samples for each model. Acceptance ratios were checked and found to be reasonable for all the traces while mixing was verified by visual inspection of the concatenated/thinned groups of chains and the auto-correlation plots. For the flux field, we considered 10 of the spatial locations, chosen at random, out of the possible 122 locations. Since $\lambda$ was well constrained by the inventory, there was no noticeable additional computational burden in sampling from its conditional distribution. The computational bottleneck for each model was the sampling of the flux field using HMC. Each study required between 5 and 7 hours of computation time and 10 Gb (1 Gb per thread) of memory to run using \texttt{R} \citep{R} and \texttt{OpenBLAS} \citep{OpenBLAS} on a computer with 64 AMD Opteron 6376 2.3 GHz processors.

In order to study model performance we considered the root-mean-squared prediction error (RMSPE) and the mean continuous rank probability score \citep[MCRPS,][]{Gneiting_2005b} that penalises both for location error in the posterior distribution and under/over-confidence. %For example, a posterior distribution centred on the true flux value with a large prediction interval will have a higher (worse) CRPS than a distribution with a small prediction interval, but a lower (better) CRPS than a distribution with a small prediction interval and large location error. 
%We complement the CRPS with \emph{violin} plots that show where the true (unobserved) flux values lie in relation to the posterior distribution at each spatial location. Such plots can be useful for quick inspection of validity (or otherwise) of the posterior inferences. 
We also carried out a simple validation study on the mole fractions, where we  diagnosed the posterior distributions at times when observations were missing, using the same diagnostics as for the flux field. Samples from the mole-fraction field, which was fully marginalised out in our sampling scheme, were obtained by using the samples from the grouped/thinned chains. Specifically, at the $i$-th iteration we sampled $\Yvec_2^{(i)}$ from $p(\Yvec_2 \mid \Zvec_2, \Yvec_1^{(i)},\thetab_2^{(i)},\tau_2^{(i)})$, where the superscript `$(i)$' indicates the $i$-th sample.

\subsection{Results} \label{sec:results}

\begin{table}[t!]
\caption{Flux field and mole-fraction prediction diagnostics for Models 1--6, Model $1^{*}$, and Model $1^{**}$: Entries give the root-mean-squared prediction error (RMSPE) and the mean continuous rank probability score (MCRPS). Flux-field diagnostics are averages of scores at each $\svec \in D^L_1$. Mole-fraction-field diagnostics are averages of scores  at the space-time points at which observations are missing.  Model $1^{*}$ and Model $1^{**}$ denote the cases when an inventory from mainland Europe and an inventory from Northern Australia are used for $\Wvec_1$, respectively.\vspace{0.2in}}\label{tab:CRPS}

% & \multicolumn{3}{c|}{Flux (g s$^{-1}$)} & \multicolumn{3}{c|}{Mole Fraction (ppb)}  \\ \hline 
\centering
\def\arraystretch{1.2}%  

\begin{tabular}{cccccc}
  \hline
&\multicolumn{2}{c}{Flux (g s$^{-1}$)} && \multicolumn{2}{c}{Mole Fraction (ppb)}  \\ \cmidrule{2-3}\cmidrule{5-6} 
Model & RMSPE & MCRPS &~&  RMSPE & MCRPS \\ 
  \hline
  1 & 294.8 & 154.7 && 18.4 & 10.9 \\ 
  2 & 293.4 & 154.1 && 18.4 & 10.8 \\ 
  3 & 327.4 & 180.2 && 18.5 & 10.9 \\ 
  4 & 306.4 & 155.5 && 18.5 & 10.9 \\ 
  5 & 319.4 & 160.7 && 18.5 & 10.9 \\ 
  6 & 332.5 & 178.2 && 18.5 & 10.9 \\ 
  $1^{*}$  & 308.8 & 165.6 && 18.4 & 10.8 \\ 
  $1^{**}$ & 329.1 & 191.4 && 18.4 & 10.8 \\  
   \hline
\end{tabular}
\end{table}

\begin{figure}[!t]
\begin{center}
\includegraphics[width=6.0in]{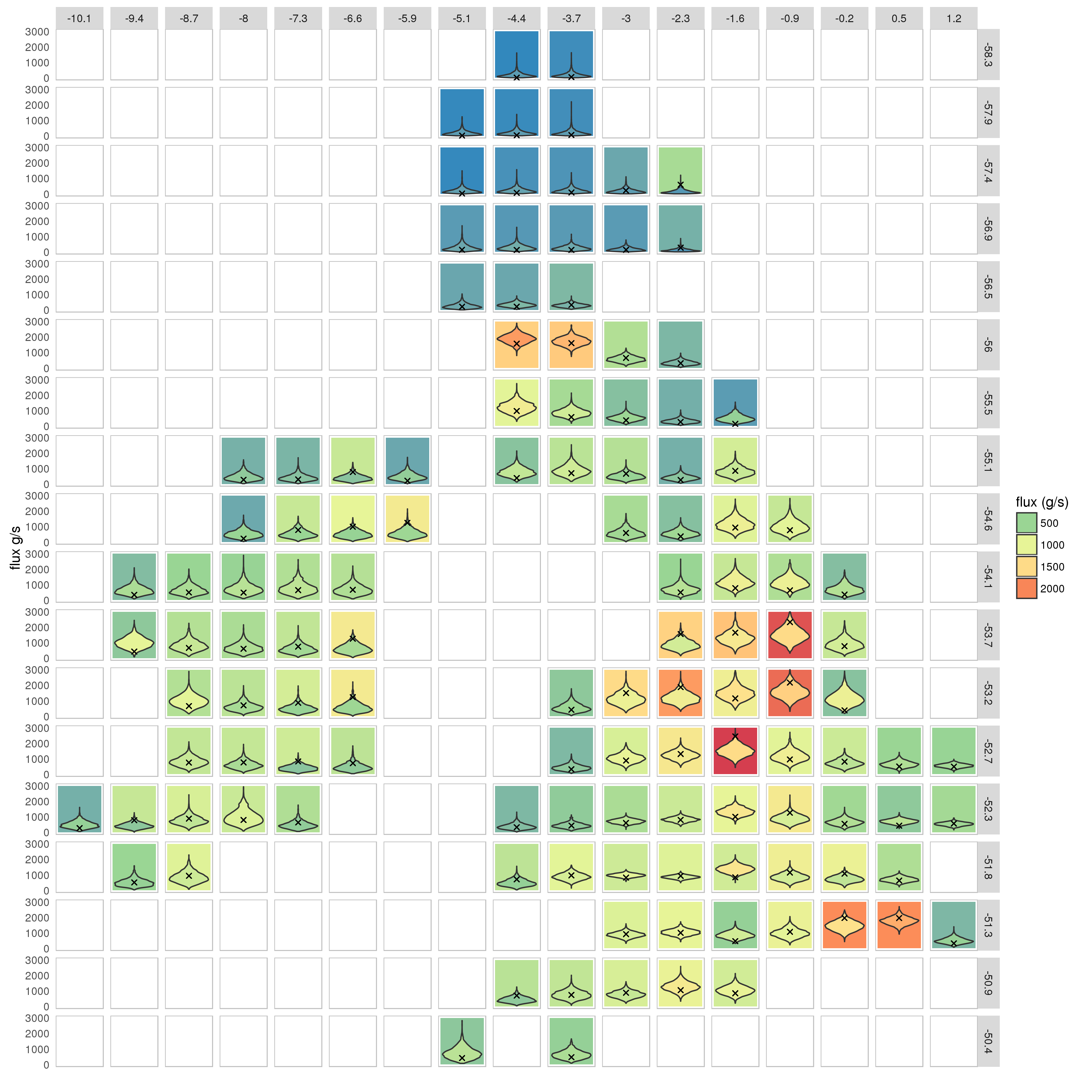}  
	\caption{Violin plot summarising the posterior distributions of the fluxes (in g/s) at each grid cell contained in $D^L_1$ for Model 1. Each violin is composed of a double-sided density plot obtained from the samples of $(\Yvec_1 \mid \Zvec_2, \Wvec_1)$. In each grid cell the `x' symbol and the background cell colour denote the true (unobserved) total flux at that grid cell. The fill colour of the density denotes the posterior median. The top and right labels on the region denote the cell centre's longitude and latitude coordinate, respectively. } \label{fig:violin}
\end{center}
\end{figure}

{\bf Flux prediction:} Flux diagnostics in Table \ref{tab:CRPS} indicate that there is no practical difference between the lognormal model (i.e., $\lambda = 0$) and the Box--Cox model when carrying out trace-gas inversion in the UK and Ireland; this is not surprising since both the posterior mean and posterior median of $\lambda$ in Model 1 is 0.19. %A paired t-test failed the reject the null hypothesis that the fourth root of the MSPE between Model 1 and Model 2 are different at a 5\% significance level, while 
In separate simulation studies on the example illustrated in \ref{sec:cum_example}, we found very little difference in the prediction performance between the lognormal process and the Box--Cox process at these small values for $\lambda$. 
%The slight lack of performance in this case is attributed to unmodelled heterogeneity in the flux field. In Fig.~\ref{fig:error_maps} we show the regions where Model 1 outperforms Model 3 (in terms of absolute error with respect to the posterior mean) and the regions where Model 2 outperforms Model 1. These regions, which are similar (78\% of the grid cells coincide), correspond to areas where inventory fluxes are either relatively low or relatively high and thus easier to fit with lower values of $\lambda$. Clearly, the Box--Cox model compensates for this heterogeneity, resulting in lower prediction errors in the regions where fluxes are not at one extreme end of the scale or another, but higher errors otherwise. 
On the other hand, as seen from Table \ref{tab:CRPS}, the truncated Gaussian model (Model 3) performs considerably worse than Models 1 and 2. %This is partially a consequence of the posterior distributions $\{(g(Y_{1,i}) \mid \Zvec_2,\Wvec_1) : i = 1,\dots,|D_1^L|\}$ containing appreciable probability mass close to $-1$, at which point $g(Y_{1,i})$ is truncated. Hence, in this application, for $\lambda = 1$ the Box--Cox transformation does not yield a distribution that is approximately multivariate normal; that is, the assumed model is mis-specified, thus contributing to poor predictive performance.  
Further, the spatially uncorrelated Box--Cox and lognormal models perform considerably worse than their correlated counterparts, both in terms of RMSPE and CRPS. %This result was expected given the nature of the data which clearly is spatially correlated. 

% \begin{figure}[!t]
% \begin{center}
% \includegraphics[width=3.1in]{./error_map1.png}  
% \includegraphics[width=3.1in]{./error_map2.png}  
% 	\caption{Left panel: Regions where Model 1 outperforms Model 3 (denoted by 1). Right panel: Regions where Model 2 outperforms Model 1 (denoted by 1). A model outperforms another model in a grid cell if the absolute error (with respect to the posterior mean) in that grid cell is less than that of the other model.} \label{fig:error_maps}
% \end{center}
% \end{figure}

In Fig.~\ref{fig:violin}, we provide a violin plot \citep{Hintze_1998}, which shows where the true flux values (here, the UK and Ireland inventory) lie in relation to the posterior distribution of the flux at each spatial location for the Box--Cox process (Model 1). The colour of the violin is related to the posterior median, while the background colour of each grid cell denotes the true flux. Most of the time, the true (in this case the inventory) flux values coincide with regions of high posterior probability density. Although the width of the intervals containing high posterior density are large for some regions, the colour map indicates a broad agreement between the posterior median and the true flux. We can see however that the extremes, namely the low emissions in the Scottish Highlands and the high emissions in the English Midlands are not correctly captured. The lognormal process (Model 2) fared slightly better in this respect, while the poor diagnostics associated with the truncated Gaussian model (Models 3 and 6) are attributed to over- and under-estimating the flux in these regions, respectively.

\begin{figure}[!t]
\begin{center}
\includegraphics[width=6.0in]{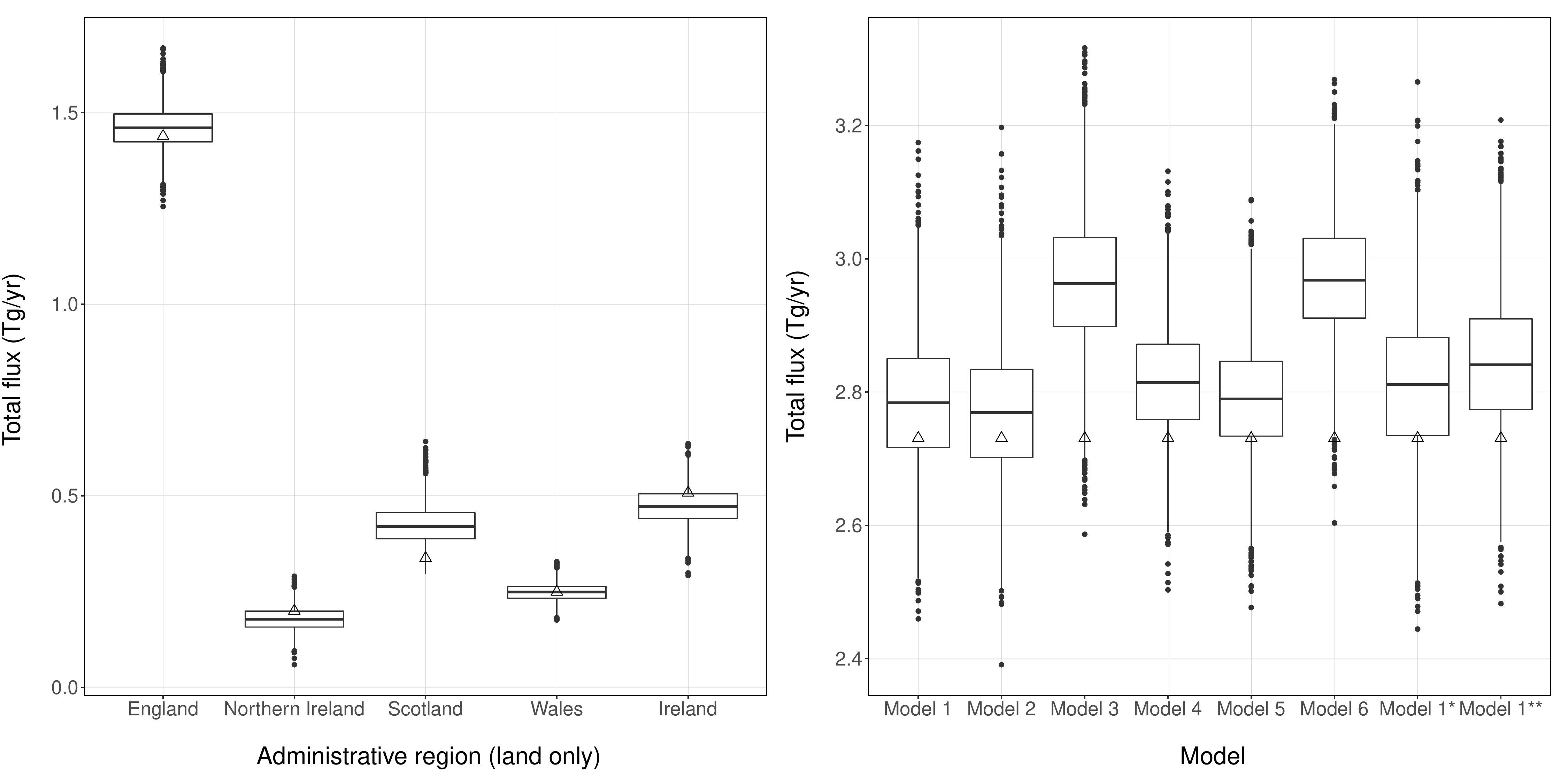}  
	\caption{Box-and-whisker plots summarising the posterior distribution of country-level and whole-domain flux aggregates. Left panel: The total flux in the land territories of England, Northern Ireland, Scotland, Wales, and Ireland for Model 1. The triangle denotes the true (unobserved) flux. Right panel: The total flux in the land territories of the UK and Ireland for Models 1--6, Model $1^{*}$, and Model $1^{**}$. The triangle denotes the true (unobserved) total flux of 2.73 Tg yr$^{-1}$. The boxes denote the interquartile range, the whiskers extend to the last values that are within 1.5 times the interquartile range from the quartiles, and the dots show the samples that lie beyond the end of the whiskers.} \label{fig:tot_flux}
\end{center}
\end{figure}

Although the uncertainty on the flux in the \emph{individual} grid cells is quite large (as can be seen from Fig.~\ref{fig:violin}), the uncertainty on the \emph{sum} of fluxes is not so poorly constrained. This is a natural consequence of averaging and of the linear operator obtained from the transport model, which integrates the flux over a large domain at each time point. We consider two levels of aggregation that are important, the entire geographic region (i.e., the UK and Ireland together), and the smaller, country level of aggregation (England, Northern Ireland, Scotland, Wales, and Ireland separately). Total flux for land territories at the country level for Model 1 are given in Fig.~\ref{fig:tot_flux}, left panel. These distributions are well constrained and, with the exception of Scotland, contain the true flux well within the 95\% credible interval. The poorer performance in Scotland is due to the near-zero fluxes in the Highlands that take anomalously low values. Clearly, inferences on fluxes become more well constrained when spatially aggregating to the resolution of countries within the UK and Ireland.

The posterior distribution of total methane emissions in the entire geographic region are summarised in Fig.~\ref{fig:tot_flux}, right panel. The truncated Gaussian models (i.e., Models 3 and 6) do not contain the true value (2.73 Tg yr$^{-1}$) within the 95\% credible interval. Interestingly, Model 1* (that uses the mainland Europe inventory) performed just as well as Model 1. Clearly, the sensitivity to the inventory when inferring over aggregated regions is less than when predicting at a high resolution, as expected. From Fig.~\ref{fig:tot_flux}, right panel, it is apparent that capturing non-Gaussianity is more important than capturing spatial correlations when predicting aggregates. On the other hand, models that capture spatial correlation provide considerably better point-level predictions (compare with Table \ref{tab:CRPS}).

%An important observation for practitioners here is that in terms of total fluxes, Model $1^{*}$ (that uses the mainland Europe inventory) performed just as well as Model 1 (that included the `perfect' inventory), despite the total flux in its inventory being 1.6 Tg yr$^{-1}$. A standard data assimilation scheme would likely have produced erroneous results with the use of this mis-specified inventory. From Fig.~\ref{fig:tot_flux}, right panel, it is apparent that capturing non-Gaussianity is more important than capturing spatial correlations when predicting aggregates, although models that capture spatial correlation provide considerably better pointwise predictions (see Table \ref{tab:CRPS}).

{\bf Mole-fraction-field prediction:} We carry out inference on the mole-fraction field at the space-time points for which no mole-fraction observation is available (note that $b_t(\svec,\cdot)$ is still evaluated at these space-time points and, hence, the true mole fraction is known). From Table \ref{tab:CRPS}, we see that all methods have similar out-of-sample mole-fraction prediction performance. This is due to the ill-posed nature of the problem: There are a large number of plausible flux fields that yield similar observations. Hence, unless $b_t(\svec,\cdot)$ evaluated at the validation space-time points is very different from all other evaluations (which is unlikely due to prevailing winds and recurring meteorology patterns) mole-fraction prediction performance clearly cannot be used to draw any conclusions on the flux-field prediction performance.

\begin{figure}[!t]
\begin{center}
\includegraphics[width=6.0in]{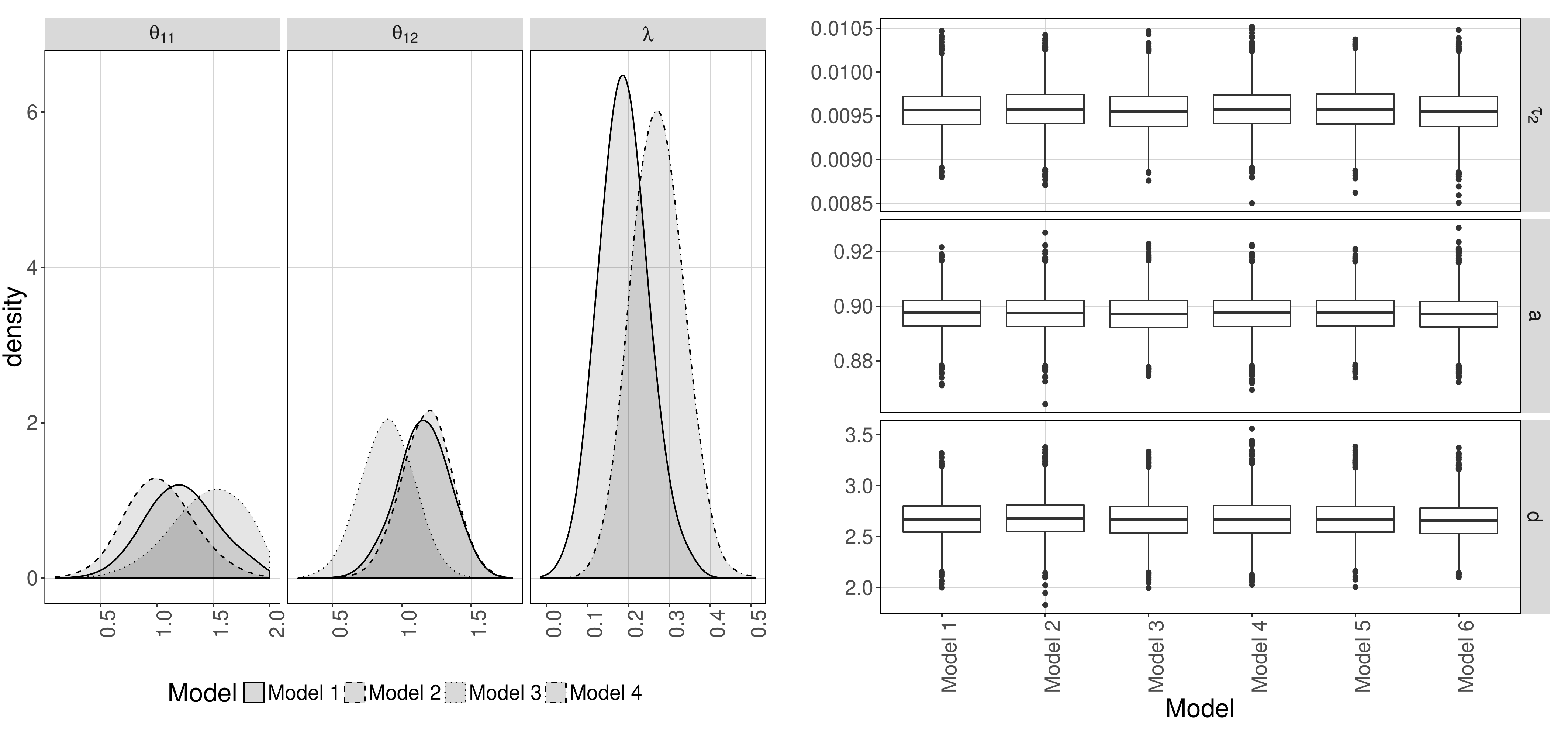}  
	\caption{Left panel: Posterior densities of the parameters $\theta_{11}, \theta_{12}$, and $\lambda$, from left to right, which appear in the flux field for Models 1--4. Note that the spatial correlation function is specified to be the indicator function in Models 4--6. Since the transformation is also assumed known for Models 5--6, no flux-field parameters need to be inferred for these models.  Right panel: Box-and-whisker plots of samples from the marginal posterior distributions of the parameters  $\tau_2, a$, and $d$, from top to bottom, which appear in the mole-fraction discrepancy term for Models 1--6. See Fig.~\ref{fig:tot_flux} for a description of the box-and-whisker plot.} \label{fig:theta_post}
\end{center}
\end{figure}

{\bf Posterior distributions of parameters:} Flux-parameter posterior distributions for Models 1--6 are summarised in Fig.~\ref{fig:theta_post}, left panel, while mole-fraction-parameter posterior distributions are summarised in Fig.~\ref{fig:theta_post}, right panel (recall that the true values are $\tau_2 = 0.01$ ppb$^{-2}, a = 0.9$, and $d = 2.5^{\circ}$). Irrespective of the model used to describe the flux field, posterior inferences on the parameters describing the discrepancy term $\zeta_t(\cdot)$ were practically identical for each case. This is reassuring, as it indicates that inferences over the discrepancy $\{\zetab_t : t \in \mathcal{T}\}$ are well constrained and, by implication, that inferences over $\{\Bmat_t\Yvec_1 : t \in \mathcal{T}\}$ are as well. This does not imply that inferences over $\Yvec_1$ are well constrained, and posterior marginal uncertainty over the elements of $\Yvec_1$ clearly depends on the dimensionality of $\Yvec_1$ due to the aggregation induced by the matrices $\{\Bmat_t : t \in \mathcal{T}\}$. 

{\bf Inventory sensitivity: }Over-reliance on the quality of the `prior inventory,' and uncertainties associated with it, is an ongoing topic of concern in atmospheric trace-gas inversion \citep[e.g.,][]{Berchet_2015}. In Table \ref{tab:CRPS}, we give diagnostics for two other models (Model $1^{*}$ and Model $1^{**}$) that are identical to Model 1 but with inventories $\Wvec_1^*$ (taken from mainland Europe containing parts of France, Switzerland, and Germany) and $\Wvec_1^{**}$ (taken from Northern Australia) instead. The inventory $\Wvec_1^*$ is, at first sight, entirely different from that attributed to the UK and Ireland (see Fig.~\ref{fig:all_inventories}). However, it contains similar spatial properties that aid in the assimilation; in particular, the maximum likelihood estimate of $\lambda$ obtained using \texttt{geoR} \citep{geoR}  with $\Wvec_1^*$ as data, is 0.21. As seen from Table \ref{tab:CRPS}, the quality of flux prediction for Model 1$^*$ deteriorates but not drastically so.  

On the other hand, the inventory $\Wvec_1^{**}$ is much smoother (recall Fig.~\ref{fig:all_inventories}, right panel); the maximum likelihood estimate of $\lambda$ obtained using \texttt{geoR} with $\Wvec_1^{**}$ as data, is 0.83, and the posterior mean obtained using our hierarchical model with the Box--Cox process (Section \ref{sec:summary}), is 0.68. This value for $\lambda$ is considerably higher than that obtained with the correct inventory and shows the impact that the inventory has on the inference of the transformation parameter. As expected, the use of this inventory considerably deteriorates the flux prediction (see Table \ref{tab:CRPS}).

From these two experiments with $\Wvec_1^*$ and $\Wvec_1^{**}$, we conclude that while the inventory remains important in trace-gas inversion in this framework, its effect on the flux-prediction performance may not dominate. Further, all that is needed is an inventory that has similar spatial characteristics as those of the true flux field; one can justifiably expect that any accepted methane inventory for the UK and Ireland will be reasonable for use within our framework. From a statistical perspective, prior sensitivity has been shifted from the inventory (process layer) to the spatial properties of the inventory (parameter layer).

{\bf Lognormal model may not always be adequate:} From the results above, one might think that for our study there is no benefit in considering the Box--Cox trans-Gaussian model over the lognormal model. We provide a counter-example by simply replacing the inventory we used for the UK and Ireland with the one from Northern Australia (Figure \ref{fig:all_inventories}, right panel), and re-running the entire OSSE with the same meteorology and station locations as those in the UK and Ireland for Models 1--6. Since the methane emissions in this region are lower than those in the UK and Ireland, we set $\tau_2 = 4$ ppb$^{-2}$ so that the discrepancy does not overwhelm the signal of interest. Recall that the maximum likelihood estimate of $\lambda$ obtained using \texttt{geoR} for this inventory is 0.83; the posterior mean obtained for Model 1 using the true inventory in the assimilation was 0.81. From these estimates, one might expect that a lognormal spatial process model for the flux field will not perform well in this region; indeed, in Table \ref{tab:CRPS2} we see that both the Box--Cox spatial process (Model 1) and the truncated Gaussian process (Model 3) considerably outperform the lognormal process (Model 2) in this case. Again, spatially correlated models are seen to outperform the spatially uncorrelated ones in flux prediction but, more importantly, the Box--Cox model is able to adapt well in this very different scenario.

\begin{table}[t!]
\caption{Flux field and mole-fraction prediction diagnostics for Models 1--6 when the mole-fraction data is simulated from an inventory from Northern Australia: Entries give the root-mean-squared prediction error (RMSPE) and the mean continuous rank probability score (MCRPS). Flux-field diagnostics are averages of scores at each $\svec \in D^L_1$. Mole-fraction-field diagnostics are averages of scores  at the space-time points at which observations are missing.\vspace{0.2in}}\label{tab:CRPS2}

% & \multicolumn{3}{c|}{Flux (g s$^{-1}$)} & \multicolumn{3}{c|}{Mole Fraction (ppb)}  \\ \hline 
\centering
\def\arraystretch{1.2}%  

\begin{tabular}{cccccc}
  \hline
&\multicolumn{2}{c}{Flux (g s$^{-1}$)} && \multicolumn{2}{c}{Mole Fraction (ppb)}  \\ \cmidrule{2-3}\cmidrule{5-6} 
Model & RMSPE & MCRPS &~&  RMSPE & CRPS \\ 
  \hline
  1 &  13.7 & 7.1 &&  0.8 & 0.5 \\ 
  2 &  17.8 & 10.3 &&  0.9 & 0.5 \\ 
  3 &  13.9 & 7.2 &&  0.8 & 0.5 \\ 
  4 &  18.8 & 10.4 &&  0.9 & 0.5 \\ 
  5 &  24.5 & 14.3 &&  0.9 & 0.5 \\ 
  6 &  19.1 & 10.5 &&  0.9 & 0.5 \\ 
   \hline
\end{tabular}
\end{table}

Reproducible code for this OSSE is available from \url{https://github.com/andrewzm/atminv}. 

\section{Discussion}\label{sec:conc}

Lognormality in flux-field modelling was considered for methane flux inversion by \citet{Ganesan_2014} and \citet{Ganesan_2015}, however they did not make use of spatial models. Spatial lognormality of the flux field was subsequently considered in \citet{Zammit_2015}, in which a two-stage inferential approach was adopted. In this article, we extend these works by proposing a new class of non-Gaussian spatio-temporal bivariate models for use in atmospheric trace-gas inversion. The bivariate model may be fully characterised through the use of cumulant functions (\ref{sec:CCFs}). It is very flexible, and yet it requires only the specification of a univariate Box--Cox spatial process, a univariate Gaussian spatio-temporal field, and an interaction function. Most previous works in this area use models that contain only parts of the class of models we consider in this article.

In the hierarchical model of Section \ref{sec:summary}, the quantity of interest, $\Yvec_1$, is not directly observed. This implies that the mole-fraction data is \emph{conditionally uninformative} of the flux field \citep{Poirer_1998} and, in this case, the quantity $\Yvec_1$ is frequently termed \emph{unidentifiable}. Unidentifiability does not prohibit Bayesian learning on $\Yvec_1$, since $\Yvec_1$ and $\Yvec_2$ are never \emph{a priori} independent in this problem. \citet{Xie_2006} proposed a way to measure the permissible extent of learning obtained from the data in this context of ``unidentifiability.'' Such measures may prove useful for trace-gas-inversion experimental design through the use of simulation studies, such as the one conducted in this article.

We implemented a fully Bayesian approach to flux inversion, in which a degree of non-Gaussianity was incorporated through the Box--Cox-transformation parameter. Our results provide further evidence that both non-Gaussianity \emph{and} spatial correlation are particularly important features to model in trace-gas inversion of methane and, further, that the degree of non-Gaussianity is also important. The truncated Gaussian assumption \citep[adopted in][]{Miller_2014} seems unsuitable for the UK and Ireland case, while the lognormal process \citep{Zammit_2015} seems unsuitable for parts of Northern Australia. The Box--Cox process adapts well to both and more generally to largely different types of flux fields.% although we stress that this conclusion could be due to model mis-specification from assuming normality of $\widetilde{Y}_1$ (and, hence, from ignoring the normalising factor in inference). 

%From a theoretical point of view, this article considers a new way in which to construct non-Gaussian bivariate spatial models and a novel application in which they can be used.

Covariates, such as population density, could be used to describe spatial features, and it is also possible that their inclusion could motivate the use of a Gaussian field for the stochastic component that remains. However, the use of covariates does not especially simplify the problem. In this example, covariates will almost certainly need to be spatially weighted and, further, their inclusion precludes the possibility for using the inventory in the assimilation, since these inventories are constructed using surrogate information such as population density. The advantage of only using second-order and higher-order prior descriptors is that the posterior distribution of the spatial pattern of fluxes is predominantly data-driven. % Of course, a purely geostatistical approach is possible, and the bivariate modelling approach we present is easily amenable to this case.
Critically, in this study we have shown that usable flux inferences can be obtained at several resolutions of importance (e.g., the country level) with a reduced reliance on the inventory. This result may have implications for other geophysical fields of study that rely heavily on data assimilation.

Although transport models are in constant development and regularly validated \citep[e.g.,][]{Ahmadov_2009,Ryall_1998}, they are by no means perfect. We acknowledge this with the use of a discrepancy term that we model as conditionally Gaussian and space-time separable (see \eqref{eq:Y2_time}). This is overly simplistic; however, there is not enough information in the data to warrant a more complex model. For short-lived gases such as carbon monoxide, the discrepancy term also captures misspecification arising from linearisation. If nonlinearity of the flux-mole-fraction mapping is deemed important to model, then one might consider replacing the linear mapping with a higher-order one or even a stochastic one (using, for example, a Gaussian process emulator). In the latter case, computation and model interpretability are rendered more difficult. For example, the bivariate process of Section \ref{sec:nonGaussian} would then need to be interpreted using \emph{modified} cumulants  \citep{Schultz_1978}.%, for which the validation experiments mentioned earlier, in the context of assessing flux-field identifiability, play a central role.

Finally, a comment is needed on inferring the transformation parameter $\lambda$ within our framework. With both the UK/Ireland and the Australia OSSE simulations, fitting a standard Box--Cox spatial model to the inventory data using \texttt{geoR} yielded maximum likelihood estimates  within 0.02 of the posterior mean obtained using our fully  Bayesian approach (on the models with the correct inventory). Thus, as expected, the data in trace-gas inversion is not informative of $\lambda$, although clearly this parameter plays a big role in the flux-prediction performance.  \citet{Box_1964} suggested that a value of $\lambda$ be selected in light of its posterior distribution, following which a standard analysis with fixed $\lambda$ could be carried out. \citet{Christensen_2001} further argue that only a select-few `interpretable' values for $\lambda$ should be chosen and that standard maximum likelihood can guide one's choice. We do not oppose these statements and, in light of the insensitivity of the mole-fraction data on $\lambda$ in this application, one might prefer to estimate $\lambda$ offline and use it as a ``plug-in.'' This pragmatic approach is unlikely to adversely affect the results in any meaningful way in regional inversions.

\section*{Acknowledgements}
As well as their University of Wollongong appointments, Andrew Zammit-Mangion is Honorary Research Fellow at the University of Bristol, and Noel Cressie is Distinguished Visiting Scientist at the Jet Propulsion Laboratory of the National Aeronautics and Space Administration (NASA). Noel Cressie's research was partially supported by a 2015--2017 Australian Research Council Discovery Project, DP150104576, and partially supported by NASA grant NNH11ZDA001N-OCO2. Anita Ganesan was funded by the Natural Environment Research Council (NERC) Independent Research Fellowship NE/L010992/1. We would like to thank Alistair J. Manning for providing us with the NAME data used in this study.  The UK National Atmospheric Emissions Inventory (NAEI) used here was funded by DECC, the Department for Environment, Food and Rural Affairs (DEFRA), the Scottish Government, the Welsh Government, and the Northern Ireland Department of Environment. We would also like to thank Bohai Zhang for providing some references, Jonathan Rougier for providing the \emph{R} software for slice sampling, and Matt Rigby for ongoing discussions related to this application. Finally, our sincere appreciation goes to two referees and the Associate Editor for constructive comments on the initial submission.

\section*{References}

\bibliography{./Spat_bib}

\appendix

\section{Derivation of cross-cumulant functions}\label{sec:CCFs}

Since all joint cumulants associated with independent variables are zero \citep[e.g.,][Theorem 4.16]{Severini_2005}, we focus here on the linear system \eqref{eq:Y2} in which the discrepancy term is zero. 
Consider two random vectors $\Yvec_1 \in \mathbb{R}^N$ and $\Yvec_2 \in \mathbb{R}^M$ such that $\Yvec_2 = \Amat\Yvec_1$, and $\Amat$ is a real-valued $M \times N$ matrix. Let the columns of $\Amat$ be denoted as $\avec_1,\dots,\avec_N,$ and denote the characteristic functions of $\Yvec_1$ and $\Yvec_2$ as 
$$
\varphi_{\Yvec_1}(\tvec_1) = \E(\exp(\im\tvec_1'\Yvec_1)) = \int_{\mathbb{R}^N} \exp(\im\tvec_1'\Yvec_1)  p(\Yvec_1) \intd \Yvec_1,
$$
and
$$
\varphi_{\Yvec_2}(\tvec_2) = \E(\exp(\im\tvec_2'\Yvec_2)) = \int_{\mathbb{R}^M} \exp(\im\tvec_2'\Yvec_2)  p(\Yvec_2) \intd \Yvec_2,
$$
respectively. The distribution of $\Yvec_1$ is obtained from
$$ p(\Yvec_1) = \int_{\mathbb{R}^M} p(\Yvec_2 \mid \Yvec_1)p(\Yvec_1) \intd \Yvec_2 = \int_{\mathbb{R}^M} \delta(\Yvec_2 - \Amat\Yvec_1)p(\Yvec_1)\intd \Yvec_2,$$
where $\delta(\cdot)$ is the Dirac delta function. Let $\Yvec \equiv (\Yvec_1',\Yvec_2')'$ and $\tvec_\Yvec \equiv (\tvec_1',\tvec_2')'$. Then the characteristic function of $\Yvec$, $\varphi_\Yvec(\tvec_\Yvec) = \E(\exp(\iota\tvec_\Yvec'\Yvec)),$ is given by
\begin{align*}
 \varphi_\Yvec(\tvec_\Yvec) &= \int_{\mathbb{R}^N}\int_{\mathbb{R}^M} \exp\left(\im[\tvec_1', \tvec_2']\begin{bmatrix}\Yvec_1 \\ \Yvec_2 \end{bmatrix}\right)p(\Yvec_2 \mid \Yvec_1)p(\Yvec_1)\intd\Yvec_2\intd\Yvec_1 \\
&= \int_{\mathbb{R}^N} \exp\left(\im[\tvec_1', \tvec_2']\begin{bmatrix}\Yvec_1 \\ \Amat\Yvec_1 \end{bmatrix}\right)p(\Yvec_1)\intd\Yvec_1 \\
&= \int_{\mathbb{R}^N} \exp\left(\im[\tvec_1', \tvec_2']\begin{bmatrix}\Imat \\ \Amat \end{bmatrix}\Yvec_1\right)p(\Yvec_1)\intd\Yvec_1.
\end{align*}
We therefore obtain the result,
\begin{equation}\label{eq:CF_Z}
\varphi_\Yvec(\tvec_\Yvec)= \varphi_{\Yvec_1}(t_{11} + \tvec_2'\avec_1,\dots,t_{1N} + \tvec_2'\avec_N).
\end{equation}
Equation \eqref{eq:CF_Z} is a natural extension to the result of \citet[][Appendix I]{Kuznetsov_1965}, who showed that 
\begin{equation}
\varphi_{\Yvec_2}(\tvec_2)= \varphi_{\Yvec_1}(\tvec_2'\avec_1,\dots,\tvec_2'\avec_N),
\end{equation}
but \eqref{eq:CF_Z} allows us to derive \emph{all} cross-cumulant functions as limits of Riemann sums. 

Returning to the original application, consider a random field $Y_2(\svec) = \int_D b(\svec,\uvec)Y_1(\uvec) \intd \uvec$ (i.e., let $\zeta(\cdot) = 0$). If $Y_1(\cdot)$ has continuous realisations with probability one, and the function $b(\svec,\uvec)$ is continuous, then this integral can be represented approximately as,
$$ Y_2(\svec_i) = \sum_{j=1}^N Y_1(\uvec_j)b(\svec_i,\uvec_j)\Delta_{\uvec_j}; \quad \svec_i \in  D,$$
where $N$ is large and $\Delta_{\uvec_j}$ is a small region in $D$ centred at $\uvec_j$. Now, let $\Yvec_1 \equiv (Y_1(\uvec_j) : j = 1,\dots,N)'$ and  $\Yvec_2 \equiv (Y_2(\svec_i) : i = 1,\dots,M)'$. Further, recall that $\varphi_{\Yvec_1}(\tvec_1)$ and $\varphi_{\Yvec_2}(\tvec_2)$ are the characteristic functions of $\Yvec_1$ and $\Yvec_2$, respectively. We saw that $\Yvec \equiv (\Yvec_1',\Yvec_2')'$ has characteristic function $\varphi_\Yvec(\tvec_\Yvec),$ where $\tvec_\Yvec \equiv (\tvec_1',\tvec_2')'$. Then, from \eqref{eq:CF_Z}, we have the approximate representation,
\begin{equation}\label{eq:CF_Y}
\varphi_{\Yvec}(\tvec_\Yvec) = \varphi_{\Yvec_1}\left(t_{11} + \sum_{i=1}^Mt_{2i}b(\svec_i,\uvec_1)\Delta_{\uvec_1},\dots,t_{1N} + \sum_{i=1}^Mt_{2i}b(\svec_i,\uvec_N)\Delta_{\uvec_N} \right).
\end{equation}

Under conditions outlined in \citet{Kuznetsov_1965}, the characteristic function of the process $Y_1(\cdot)$ evaluated at $\{\uvec_j: j = 1,\dots,N\}$ can be expanded as a series
$$
\varphi_{\Yvec_1}(\tvec_1) = \exp\left(\im\sum_{j=1}^N \kappa_{Y_1}^1(\uvec_j)t_{1j} + \frac{\im^2}{2}\sum_{j,j' = 1}^{N} \kappa^2_{Y_1Y_1}(\uvec_j,\uvec_{j'})t_{1j}t_{1j'} + \dots\right).
$$
Using this series definition for $\varphi_{\Yvec_1}(\tvec_1)$ in \eqref{eq:CF_Y}, we obtain
\begin{align}
\varphi_{\Yvec}(\tvec_\Yvec) &= \exp\left(\im\sum_{j=1}^N \kappa_{Y_1}^1(\uvec_j)\left[t_{1j} + \sum_{i=1}^M t_{2i}b(\svec_i,\uvec_j)\Delta_{\uvec_j}\right] \right.\nonumber\\
&~~~ + \frac{\im^2}{2}\sum_{j,j' = 1}^{N} \kappa^2_{Y_1Y_1}(\uvec_j,\uvec_{j'})\left[t_{1j} + \sum_{i =1}^Mt_{2i}b(\svec_i,\uvec_j)\Delta_{\uvec_j}\right]\left[t_{1j'} + \sum_{i' =1}^Mt_{2i'}b(\svec_{i'},\uvec_{j'})\Delta_{\uvec_{j'}}\right] \nonumber\\
&~~~+ \dots\Bigg).\label{eq:CF1}
\end{align}

Now $\varphi_{\Yvec}(\tvec_\Yvec)$ may also be expanded as a series:
\begin{align}
\varphi_{\Yvec}(\tvec_\Yvec) &= \exp\left(\im\sum_{j=1}^N \kappa_{Y_1}^1(\uvec_j)t_{1j} + \im\sum_{i=1}^M \kappa_{Y_2}^1(\svec_i)t_{2i} \right.\nonumber\\
&~~~ + \frac{\im^2}{2}\sum_{i,i' = 1}^{M} \kappa_{Y_2Y_2}^2(\svec_i,\svec_i')t_{2i}t_{2i'} + \frac{\im^2}{2}\sum_{j,j' = 1}^{N} \kappa_{Y_1Y_1}^2(\uvec_j,\uvec_j')t_{1j}t_{1j'} \nonumber\\
&~~~ + \frac{\im^2}{2}\sum_{i,j' = 1}^{M,N} \kappa_{Y_2Y_1}^2(\svec_i,\uvec_j')t_{2i}t_{1j'}+ \frac{\im^2}{2}\sum_{i',j = 1}^{M,N} \kappa_{Y_1Y_2}^2(\uvec_j,\svec_i')t_{2i'}t_{1j} + \dots \Bigg).\label{eq:CF2}
\end{align}

\noindent Expanding \eqref{eq:CF1} and comparing coefficients of the products of elements in $\tvec_\Yvec$ to those in \eqref{eq:CF2}, we deduce that
\begin{align*}
\kappa_{Y_2}^1(\svec_i) &= \sum_{j=1}^Nb(\svec_i,\uvec_j)\kappa_{Y_1}^1(\uvec_j)\Delta_{\uvec_j}, \\
\kappa_{Y_1Y_2}^2(\uvec_j,\svec_i') &= \sum_{j'=1}^N\kappa_{Y_1Y_1}^2(\uvec_j,\uvec_{j'})b(\svec_i',\uvec_j')\Delta_{\uvec_j'}, \\
\kappa_{Y_2Y_2}^2(\svec_i,\svec_i') &= \sum_{j,j'=1}^Nb(\svec_i,\uvec_j)b(\svec_{i'},\uvec_{j'})\kappa_{Y_1Y_1}^2(\uvec_j,\uvec_{j'})\Delta_{\uvec_j}\Delta_{\uvec_j'}, \\
&\vdots
\end{align*}
These sums tend to Riemann integrals, such as those shown in \eqref{eq:12cumulants} and \eqref{eq:12cumulantsb}, provided $b(\cdot,\cdot)$, $\kappa_{Y_1}^1(\cdot), \kappa_{Y_1Y_1}^2(\cdot,\cdot), \dots$ are Riemann-integrable. Similar arguments can be used to derive third-order cross-cumulants. For example, it can be shown that
\begin{equation*}
\kappa^3_{Y_1Y_1Y_2}(\uvec_j,\uvec_{j'},\svec_{i''}) = \int \kappa^3_{Y_1Y_1Y_1}(\uvec_j,\uvec_{j'},\uvec_{j''})b(\svec_{i''},\uvec_{j''})\intd\uvec_{j''},
\end{equation*}
and thus we can formulate all the cross-cumulants of the joint process $(Y_1(\cdot),Y_2(\cdot))$ from those of $Y_1(\cdot)$ through the interaction function $b(\cdot,\cdot)$. 

These results can be extended to \eqref{eq:Y2}, where the discrepancy term is included (and is independent of $Y_1(\cdot)$) by taking advantage of the additive property of cumulants and the fact that all joint cumulants associated with independent variables are zero \citep[][Theorem 4.16]{Severini_2005}. 

\section{Example illustrating higher-order cross-cumulant functions for bivariate processes}\label{sec:cum_example}

From the relationship between moments and cumulants, it can be shown that $\kappa_{Y_i}^1(\cdot) = \E(Y_i(\cdot)), \kappa_{Y_iY_j}^2(\cdot,\cdot) = C_{Y_iY_j}(\cdot,\cdot) \equiv \cov(Y_i(\cdot),Y_j(\cdot))$ and $\kappa_{Y_iY_jY_k}^3(\cdot,\cdot,\cdot) = \E((Y_i(\cdot) - \mu_i(\cdot))(Y_j(\cdot) - \mu_j(\cdot))(Y_k(\cdot) - \mu_k(\cdot)))$.~These relationships hold even if the moment generating function does not exist \citep[as is the case, for example, with the lognormal process; see][p.~116]{Severini_2005}.\ifdetails\footnote{\citet{Severini_2005} states that \emph{although cumulants may be calculated directly from the characteristic function, there are relatively few cases in which the moment-generating function of a distribution does not exist, while the characteristic function is easily calculated and easy to differentiate. It is often simpler to calculate the cumulants by calculating the moments of the distribution and then using the relationship between cumulants and moments.}} \fi~Assume, for illustration, that $Y_1(\cdot)$ is a spatial lognormal process such that $\ln Y_1(\cdot)$ is a Gaussian process with mean function $\widetilde\mu_1(\cdot)$ and covariance function $\widetilde C_{Y_1Y_1}(\cdot,\cdot)$; then the first two spatial cumulant functions are the well known mean and covariance functions of the lognormal process \cite[e.g.,][]{Aitchison_1957}, while the third-order spatial cumulant function can be obtained by repeated application of the standard expression used to find the moments of a multivariate lognormal distribution \cite[e.g.,][p.~219]{Kotz_2002}:\ifdetails \footnote{Consider a random vector $\Xvec \in \mathbb{R}^3$ that is multivariate lognormal such that $\ln(\Xvec)$ is multivariate normal with mean $\muvec$ and covariance matrix $\Cmat$. Then, the third-order cumulant (first-order in each variable), $\kappa_{X_1X_2X_3}^3$, is 
$$
\kappa^3_{X_1X_2X_3} = \E(X_1X_2X_3) - \E(X_1)\E(X_2X_3) - \E(X_2)\E(X_1X_3) - \E(X_3)\E(X_1X_2) + 2\E(X_1)\E(X_2)\E(X_3).
$$
The moments of the multivariate lognormal distribution, parameterised by the order $\tvec$, are \citep[][p.~219]{Kotz_2002} 
$$
\E(\exp(\tvec' \Xvec)) = \exp\left(\tvec'\muvec + \frac{1}{2}\tvec'\Cmat\tvec\right),
$$
and therefore  
\begin{align*}
\kappa^3_{X_1X_2X_3} &= \exp(\mu_1 + \mu_2 + \mu_3 + \frac{1}{2}(C_{11} + 2C_{21} + 2C_{31} + C_{22} + 2C_{23} + C_{33})) \\
&~~~ - \exp(\mu_1 + \frac{1}{2}C_{11})\exp(\mu_2 + \mu_3 + \frac{1}{2}(C_{22} + 2C_{23} + C_{33})) \\
&~~~ - \exp(\mu_2 + \frac{1}{2}C_{22})\exp(\mu_1 + \mu_3 + \frac{1}{2}(C_{11} + 2C_{13} + C_{33})) \\
&~~~ - \exp(\mu_3 + \frac{1}{2}C_{33})\exp(\mu_1 + \mu_2 + \frac{1}{2}(C_{11} + 2C_{12} + C_{22})) \\
&~~~ + 2\kappa_{X_1}^1\kappa_{X_2}^1\kappa_{X_3}^1 \\
&= \kappa_{X_1}^1\kappa_{X_2}^1\kappa_{X_3}^1[\exp(C_{21} + C_{31} + C_{23}) - \exp(C_{23}) - \exp(C_{13}) - \exp(C_{12}) + 2]
\end{align*}
where in the last line we have used $\kappa_{X_i}^1 = \exp(\mu_i + \frac{1}{2}C_{ii})$.

} \fi
\begin{align*}
\kappa^3_{Y_1Y_1Y_1}(\uvec_1,\uvec_2,\uvec_3) = &\kappa_{Y_1}^1(\uvec_1)\kappa_{Y_1}^1(\uvec_2)\kappa_{Y_1}^1(\uvec_3)\Big(\exp(\widetilde C_{Y_1Y_1}(\uvec_1,\uvec_2) + \widetilde C_{Y_1Y_1}(\uvec_1,\uvec_3) + \widetilde C_{Y_1Y_1}(\uvec_2,\uvec_3)) \nonumber \\
&\qquad  - \exp(\widetilde C_{Y_1Y_1}(\uvec_1,\uvec_2)) - \exp(\widetilde C_{Y_1Y_1}(\uvec_1,\uvec_3)) - \exp(\widetilde C_{Y_1Y_1}(\uvec_2,\uvec_3)) + 2\Big).
\end{align*}

\begin{figure}[!t]
\begin{center}
\includegraphics[width=6.0in]{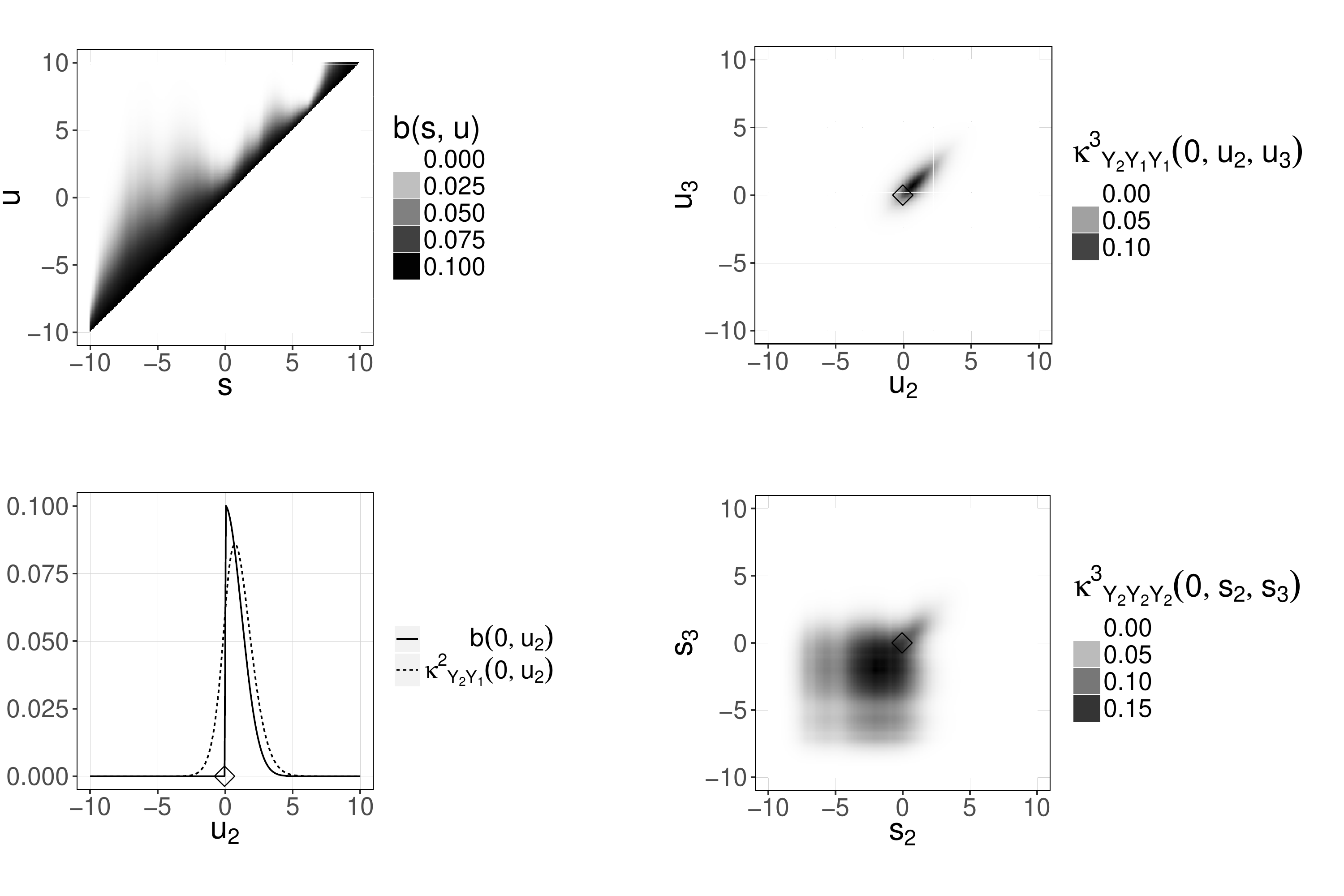}  
	\caption{Properties of a non-Gaussian process constructed from \eqref{eq:Y2} with $\zeta(\cdot) = 0$ and $Y_1(\cdot)$ a lognormal process. Top-left panel: The interaction function $b(s,u); s,u \in D^L$. Bottom-left panel: The interaction function $b(0,u_2); u_2 \in D^L$, and the second-order cross-cumulant function $\kappa^2_{Y_2Y_1}(0,u_2); u_2 \in D^L$. Top-right panel: The third-order cross-cumulant function $\kappa^3_{Y_2Y_1Y_1}(0,u_2,u_3); u_2,u_3 \in D^L$. Bottom-right panel:  The third-order auto-cumulant function $\kappa^3_{Y_2Y_2Y_2}(0,s_2,s_3); s_2, s_3 \in D^L$. In all panels, the diamond symbol denotes the origin $u_2 = u_3 = s_2 = s_3 = 0$.} \label{fig:cumulants}
\end{center}
\end{figure}

We use these expressions to illustrate second-order and third-order properties of the joint process $(Y_1(\cdot),Y_2(\cdot))$ in a simple scenario with $\zeta(s) = 0; s\in D = [-10,10] \subset\mathbb{R}$ (i.e., $d=1$). Let $Y_1(\cdot)$ be a lognormal process on $D$, with $\ln Y_1(\cdot)$ having mean function $\widetilde \mu_1(s) = -2$ and covariance function
\begin{equation}\label{eq:C11}
\widetilde C_{Y_1Y_1}(u_1,u_2 \mid \thetab_1) = \frac{1}{\tau_1}\exp\left(- \theta_{11} | u_1 - u_2 |^{\theta_{12}} \right),
\end{equation}
where $\thetab_1 \equiv (\theta_{11},\theta_{12})'$, and $\theta_{11} > 0, 0 < \theta_{12} < 2$. Here, we set the scale parameter $\theta_{11} = 0.8$, the smoothness parameter $\theta_{12} = 1.7$, and the precision parameter $\tau_1 = 1$.  We let $b(s,u); s,u \in D,$ be a truncated Gaussian density function centred and truncated at $s$ with variance varying smoothly with $u$, representing a typical highly directional interaction function in $u$; see Fig.~\ref{fig:cumulants}, top-left panel.  We now concentrate on the spatial location $s =0$, and we carry out the integrations required to find the spatial cumulant functions, such as those in \eqref{eq:12cumulants}--\eqref{eq:cross-cum}, on a discrete grid $D^L = \{-9.9, -9.7,\dots, 9.9\}$. In the bottom-left, top-right, and bottom-right panels of Fig.~\ref{fig:cumulants}, we show the second-order cross-cumulant function (i.e., cross-covariance function) $\kappa^2_{Y_2Y_1}(0,u_2)$, the third-order cross-cumulant function $\kappa^3_{Y_2Y_1Y_1}(0,u_2,u_3)$, and the third-order auto-cumulant function $\kappa^3_{Y_2Y_2Y_2}(0,s_2,s_3)$, respectively, for $u_2,u_3,s_2, s_3 \in D^L$. It is clear from these panels  that asymmetry is naturally present in the cross-covariance functions and the third-order cross-cumulant functions. Also, the two third-order cross-cumulant functions are markedly different from each other; in contrast, if $Y_1(\cdot)$ were Gaussian the third-order cumulants would have been zero everywhere. Of course, $Y_1(\cdot)$ need not be lognormal, but the spatial cumulant functions of $Y_2(\cdot)$ and those of the joint process $(Y_1(\cdot),Y_2(\cdot))$ depend critically on the spatial cumulant functions of $Y_1(\cdot)$. 

%There are several approaches to modelling non-Gaussian random fields such as $Y_2(\cdot)$ \citep[see][for a review]{Allard_2012}. The attractive notion of the conditional approach to constructing such fields is that one need only specify one such model for $Y_1(\cdot)$, a linear operator, and a second random field $\zeta(\cdot)$. Existence of $Y_2(\cdot)$ is guaranteed under mild conditions and its properties follow directly from these three ingredients. As we show next, the ensuing non-Gaussian bivariate model is also a natural one to use in the problem of trace-gas inversion.

%We therefore propose that a flexible non-Gaussian model be used for $Y_1(\cdot)$. As we show next, the resulting non-Gaussian bivariate model is a natural choice in the problem of trace-gas inversion.

\section{Derivation of conditional distributions}\label{sec:CondDist}

\subsection{The discrepancy parameters}

Consider the conditional distribution,  $p(\tau_2, \thetab_2 \mid \Zvec_2, \Yvec_1) \propto p(\Zvec_2 \mid \tau_2,\thetab_2,\Yvec_1)p(\tau_2)p(\thetab_2)$. This is given by
\begin{align}
p(\tau_2,\thetab_2 \mid \Zvec_2,\Yvec_1) &\propto \left(\int p(\Zvec_2 \mid \Yvec_2) p(\Yvec_2 \mid \tau_2,\thetab_2,\Yvec_1)\intd \Yvec_2\right)p(\tau_2)p(\thetab_2) \nonumber \\
&\propto |\Qmat_{\zeta;\tau_2,\thetab_2}|^{1/2}\left[\int \exp\left(-\frac{1}{2}(\Zvec_2 - \Cmat\Yvec_2)'\Vmat^{-1}(\Zvec_2 - \Cmat\Yvec_2) \right.\right. \nonumber \\
&\qquad\qquad\qquad \left.\left.- \frac{1}{2}(\Yvec_2 - \Bmat\Yvec_1)'\Qmat_{\zeta;\tau_2,\thetab_2}(\Yvec_2 - \Bmat\Yvec_1)\right)\intd\Yvec_2\right]p(\tau_2)p(\thetab_2) \nonumber \\
&\propto \frac{p(\tau_2)p(\thetab_2)|\Qmat_{\zeta;\tau_2,\thetab_2}|^{1/2}}{|\Cmat'\Vmat^{-1}\Cmat + \Qmat_{\zeta;\tau_2,\thetab_2}|^{1/2}}  \label{eq:expand_exponent}\\ \nonumber
& \quad \times \exp\left(-\frac{1}{2}\Yvec_1'\Bmat'\Qmat_{\zeta;\tau_2,\thetab_2}\Bmat\Yvec_1 + \frac{1}{2}\Dvec_{\thetab_2}'(\Cmat'\Vmat^{-1}\Cmat + \Qmat_{\zeta;\tau_2,\thetab_2})^{-1}\Dvec_{\thetab_2}\right),
\end{align}
where $\Qmat_{\zeta;\tau_2,\thetab_2} \equiv \bSigma_{\zeta;\tau_2,\thetab_2}^{-1}$, $\Cmat \equiv \textrm{bdiag}(\{\Cmat_t : t \in \mathcal{T}\})$, $\Vmat \equiv \textrm{bdiag}(\{\Vmat_t : t \in \mathcal{T}\})$, $\Bmat \equiv (\Bmat_t' : t \in \mathcal{T})'$ and
$$
\Dvec_{\thetab_2} \equiv (\Cmat'\Vmat^{-1}\Zvec_2 + \Qmat_{\zeta;\tau_2,\thetab_2}\Bmat\Yvec_1).
$$
The prior distributions $p(\tau_2)$ and $p(\thetab_2)$ are bounded uniform distributions, with limits as described in Section \ref{sec:semi-empirical}.

\subsection{The flux field}

It is straightforward to show that using the \citet{Pericchi_1981} prior distribution under the assumption that the transformed field is multivariate normal, $p(\betab,\tau_1 \mid \underline\Yvec_{1}, \thetab_1,\lambda)$ is multivariate Normal-Gamma \citep[][Chapter 9]{OHagan_2000}; that is,
\begin{equation} 
(\betab,\tau_1 \mid \underline\Yvec_{1}, \thetab_1,\lambda) \sim \mathcal{N}\mathcal{G}a\left(\hat\betab_{\thetab_1,\lambda},~\underline\Xvec'\underline{\widetilde\Rmat}_{Y_1Y_1;\thetab_1}^{-1}\underline\Xvec,~|D_1^L|,~\frac{S^2_{\thetab_1,\lambda}}{2}\right),
\end{equation}
where $\mathcal{N}\mathcal{G}a(\muvec,\Qmat,\alpha_1,\alpha_2)$ denotes the multivariate Normal-Gamma distribution with location $\muvec$, scale $\Qmat$, shape $\alpha_1$ and rate $\alpha_2$.\ifdetails\footnote{If $\Xvec \sim \mathcal{NG}a(\muvec,\Qmat,\alpha_1,\alpha_2)$, then
$$
p(\Xvec) = (2\pi)^{-k/2}|\tau\Qmat|^{1/2}\frac{\alpha_2^{\alpha_1}}{\Gamma(\alpha_1)}\tau^{\alpha_1-1}\exp(-\alpha_2\tau)\exp\left(-\frac{\tau}{2}(\Xvec - \muvec)'\Qmat(\Xvec - \muvec)\right).
$$} \fi~Writing out
\begin{equation}\label{eq:margin_Bayes}
p(\underline\Yvec_{1} \mid \thetab_1, \lambda) = \frac{p(\underline\Yvec_{1} \mid \betab ,\tau_1,\thetab_1,\lambda)p(\betab,\tau_1,\thetab_1 \mid \lambda)p(\lambda)}{p(\betab,\tau_1 \mid \underline\Yvec_{1},\thetab_1,\lambda)p(\thetab_1,\lambda)},
\end{equation}
we deduce that\ifdetails\footnote{To get the result simply plug in the likelihood and the multivariate normal Gamma distribution into \eqref{eq:Y1a_cond}. When doing this we end up with exponential functions of $\tau$ and a function of $\underline\Yvec_1$ which we must ignore because the left-hand-side is independent of $\tau$. We are hence using the fact that if $f_1(x) = f_2(x)e^{y\cdot f_3(x)}$ for all $y \in Dom(y)$, then $f_1(x) = f_2(x)$.} \fi 
\begin{equation}\label{eq:Y1a_cond}
p(\underline\Yvec_1 \mid \thetab_1, \lambda) \propto \left(\frac{S^2_{\thetab_1,\lambda}}{2}\right)^{-|D^L_1|}\underline J_{\lambda}.
\end{equation}
Note that $(\underline\Yvec_1 \mid \thetab_1, \lambda)$ depends on $\underline\Yvec_1$ through $S^2_{\thetab_1,\lambda}$ as in normal-linear models \citep[][Section 9.29]{OHagan_2000} and, in our case, also on the Jacobian, $\underline J_{\lambda}$.

Now, the required full conditional distribution is
\begin{equation} \label{eq:Y1_cond4}
p(\Yvec_{1} \mid \Zvec_2, \Wvec_1,\thetab_2,\tau_2,\thetab_1,\lambda) \propto p(\Zvec_2 \mid \tau_2,\thetab_2,\Yvec_1)p(\underline\Yvec_1 \mid \thetab_1,\lambda).
\end{equation}
The second term in the product on the right-hand side of \eqref{eq:Y1_cond4} is given by \eqref{eq:Y1a_cond}, while the first term is obtained through marginalisation of $\Yvec_2,$ as in \eqref{eq:expand_exponent}. Hence,
\begin{align}
\ln p(\Yvec_{1} \mid \Zvec_2,& \Wvec_1,\thetab_2,\tau_2,\thetab_1,\lambda) \nonumber \\
&= \textrm{const} - \frac{1}{2}\Yvec_1'\Bmat'(\Qmat_{\zeta;\tau_2,\thetab_2} - \Qmat_{\zeta;\tau_2,\thetab_2}(\Cmat'\Vmat^{-1}\Cmat + \Qmat_{\zeta;\tau_2,\thetab_2})^{-1}\Qmat_{\zeta;\tau_2,\thetab_2})\Bmat\Yvec_1 \nonumber \\
&\quad+ \Yvec_1'\Bmat'\Qmat_{\zeta;\tau_2,\thetab_2}(\Cmat'\Vmat^{-1}\Cmat + \Qmat_{\zeta;\tau_2,\thetab_2})^{-1}\Cmat'\Vmat^{-1}\Zvec_2 \nonumber \\
&\quad- |D^L_1|\ln\left(\frac{S^2_{\thetab_1,\lambda}}{2}\right)+\ln \underline J_{\lambda},\label{eq:Y1_logcond}
\end{align}
where `const' collects all the terms that are not a function of $\Yvec_1$. 

For Hamiltonian Monte Carlo, we also need the derivative of \eqref{eq:Y1_logcond}. By substituting \eqref{eq:betaGLS} into \eqref{eq:S2} we can re-write $S^2_{\thetab_1,\lambda}$ as $S^2_{\thetab_1,\lambda} = \underline\Gmat_{\lambda}'\Psib_{\thetab_1}\underline\Gmat_{\lambda}$, where
$$
\Psib_{\thetab_1} = \underline{\widetilde\Rmat}_{Y_1Y_1;\thetab_1}^{-1} - \underline{\widetilde\Rmat}_{Y_1Y_1;\thetab_1}^{-1}\underline\Xvec(\underline\Xvec'\underline{\widetilde\Rmat}_{Y_1Y_1;\thetab_1}^{-1}\underline\Xvec)^{-1}\underline\Xvec'\underline{\widetilde\Rmat}_{Y_1Y_1;\thetab_1}^{-1}.
$$
Then, the required derivative is 
\begin{align*}
\frac{\partial \ln p(\Yvec_{1} \mid \Zvec_2, \Wvec_1,\thetab_2,\tau_2,\thetab_1,\lambda)}{\partial \Yvec_1'} &= -\Bmat'(\Qmat_{\zeta;\tau_2,\thetab_2} - \Qmat_{\zeta;\tau_2,\thetab_2}(\Cmat'\Vmat^{-1}\Cmat + \Qmat_{\zeta;\tau_2,\thetab_2})^{-1}\Qmat_{\zeta;\tau_2,\thetab_2})\Bmat\Yvec_1 \\
& \quad + \Bmat'\Qmat_{\zeta;\tau_2,\thetab_2}(\Cmat'\Vmat^{-1}\Cmat + \Qmat_{\zeta;\tau_2,\thetab_2})^{-1}\Cmat'\Vmat^{-1}\Zvec_2 \\
& \quad - \frac{2|D^L_1|}{S^2_{\thetab_1,\lambda}}\frac{\partial \underline\Gmat_{\lambda}'}{\partial \Yvec_{1}'}\Psib_{\thetab_1}\underline\Gmat_{\lambda} + \Gmat_{\lambda,1}^{(2)} \oslash \Gmat_{\lambda,1}^{(1)},
\end{align*}
where $\Gmat_{\lambda,1}^{(1)} \equiv (\partial g_\lambda(Y_{1,i}) / \partial Y_{1,i} : i = 1,\dots, |D^L_1|)$, $\Gmat_{\lambda,1}^{(2)} \equiv (\partial^2 g_\lambda(Y_{1,i}) / \partial Y_{1,i}^2 : i = 1,\dots, |D^L_1|)$, the operator $\oslash$ denotes element-wise division,
\begin{equation}
\frac{\partial \underline\Gmat_{\lambda}'}{\partial \Yvec_{1}'}  = \left(\frac{\partial \Gmat_{\lambda_1}'}{\partial \Yvec_1'} \quad \zerob \right) = \left(\textrm{diag}(\Gmat^{(1)}_{\lambda,1}) \quad \zerob \right),
\end{equation}
and we have used the fact that 
\begin{equation}
\frac{\partial S^2_{\thetab_1,\lambda}}{\partial \Yvec_1'} = \frac{\partial\underline\Gvec_{\lambda}'\Psib_{\thetab_1}\underline\Gvec_{\lambda}}{\partial \Yvec_1'} = 2 \left(\frac{\partial \underline\Gmat_{\lambda}'}{\partial \Yvec_{1}'}\right) \Psib_{\thetab_1} \underline\Gmat_{\lambda}.
\end{equation}\ifdetails\footnote{Consider a simple $2\times 2$ system. Then, using numerator layout notation, we have that $\frac{\partial}{\partial\Yvec_1'}g(\Yvec_1)' = \frac{\partial}{\partial\Yvec_1}g(\Yvec_1)$ and therefore
\begin{align*}
&\frac{\partial}{\partial\Yvec_1'} \begin{bmatrix} g(Y_{11}) &g(Y_{12}) \end{bmatrix}\Amat \begin{bmatrix} g(Y_{11}) \\g(Y_{12}) \end{bmatrix} \\
= &2\begin{bmatrix} \frac{\partial g(Y_{11})}{\partial Y_{11}} & \frac{\partial g(Y_{12})}{\partial Y_{11}} \\ \frac{\partial g(Y_{11})}{\partial Y_{12}} & \frac{\partial g(Y_{12})}{\partial Y_{12}}  \end{bmatrix} \Amat \begin{bmatrix} g(Y_{11}) \\g(Y_{12}) \end{bmatrix},
\end{align*}
giving the required result.
} \fi
\subsection{The flux-field parameters}

From \eqref{eq:margin_Bayes}, we have 
$$
p(\thetab_1, \lambda \mid \underline\Yvec_1) = \frac{p(\underline\Yvec_1 \mid \betab ,\tau_1,\thetab_1,\lambda)}{p(\betab,\tau_1 \mid \underline\Yvec_1,\thetab_1,\lambda)}\frac{p(\betab,\tau_1,\thetab_1\mid\lambda)p(\lambda)}{p(\underline\Yvec_1)},
$$
from which we deduce that
\begin{equation}\label{eq:joint_thetab2}
p(\thetab_1,\lambda \mid \underline\Yvec_1) \propto |\underline{\widetilde\Rmat}_{Y_1Y_1;\thetab_1}|^{-1/2} |\underline\Xmat'\underline{\widetilde\Rmat}_{Y_1Y_1;\thetab_1}^{-1}\underline\Xmat|^{-1/2}(S^2_{\thetab_1,\lambda})^{-|D^L_1|}\underline J_{\lambda}p(\thetab_1)p(\lambda).
\end{equation}
This is similar to Equation 8 in \citet{deOliveira_1997}; however they used the prior distribution given by \citet{Box_1964} instead, which results in different powers of $S^2_{\thetab_1,\lambda}$ and $\underline J_\lambda$ in \eqref{eq:joint_thetab2}.

\ifdetails
\newpage
\theendnotes
\fi

\end{document}